\RequirePackage[mathlines]{lineno} 
\documentclass[a4paper,11pt]{article}

\usepackage{jheppub} 

\usepackage[T1]{fontenc} 
\usepackage{besphysics}

\usepackage{threeparttable}
\usepackage{multirow}
\usepackage{multicol}
\usepackage{subfigure}
\usepackage{amssymb,bm,mathrsfs,bbm,amscd}
\usepackage{float}
\let\oldequation\equation
\let\oldendequation\endequation
\renewenvironment{equation}
  {\linenomathNonumbers\oldequation}
  {\oldendequation\endlinenomath}

\newcommand{\sigmode}{\Lambda_{c}^{+}\rightarrow n K^0_{\mathrm{S}} \pi^+}

\def \ee   {e^{+}e^{-}}
\def \piz  {\pi^{0}}
\def \pip  {\pi^{+}}
\def \pim  {\pi^{-}}

\def \ifb  {\mbox{fb$^{-1}$}}
\def \gev  {\mbox{GeV}}
\def \gevc {\mbox{GeV/$c$}}
\def \gevcc{\mbox{GeV/$c^2$}}
\def \mev  {\mbox{MeV}}

\def \romanOne   {\uppercase\expandafter{\romannumeral1}}
\def \romanTwo   {\uppercase\expandafter{\romannumeral2}}
\def \romanThree {\uppercase\expandafter{\romannumeral3}}
\def \romanFour  {\uppercase\expandafter{\romannumeral4}}
\def \romanFive  {\uppercase\expandafter{\romannumeral5}}
\def \romanSix   {\uppercase\expandafter{\romannumeral6}}
\def \romanSeven {\uppercase\expandafter{\romannumeral7}}
\def \romanEight {\uppercase\expandafter{\romannumeral8}}

\def \mbc {M_{\rm{BC}}}
\def \dE {\Delta E}
\def \ebeam {E_{\mathrm{beam}}}
\def \Ks {K^0_{\mathrm{S}}}

\def \lcp {\Lambda_{c}^{+}}
\def \lcm {\bar{\Lambda}_{c}^{-}}

\newcommand{\BESIIIorcid}[1]{\href{https://orcid.org/#1}{\hspace*{0.1em}\raisebox{-0.45ex}{\includegraphics[width=1em]{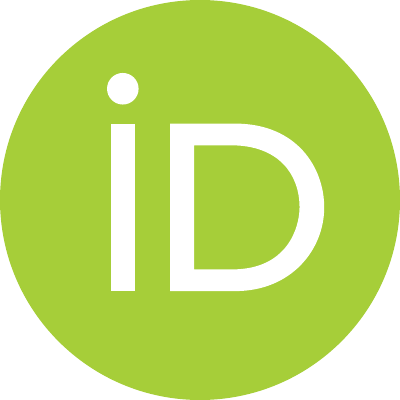}}}}

\begin{document}

\title{\boldmath Amplitude analysis of the $\Lambda_c^+\to n K^0_{\mathrm{S}} \pi^+$ decay}

\collaborationImg{\includegraphics[width=0.15\textwidth, angle=90]{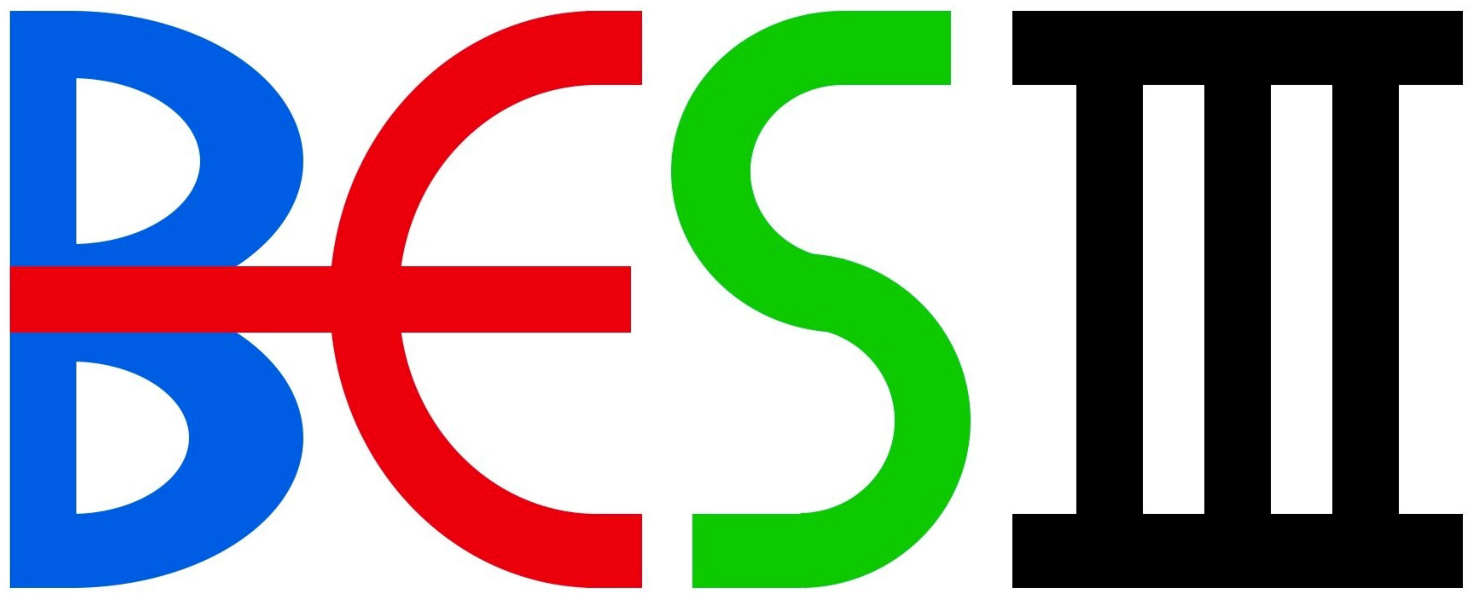}}
\collaboration{The BESIII collaboration}
\emailAdd{besiii-publications@ihep.ac.cn}

\abstract{We perform the first amplitude analysis of the charmed baryon hadronic decay $\sigmode$. The analysis uses $e^+ e^-$ annihilation data corresponding to a total integrated luminosity of 4.5~$\rm{fb}^{-1}$ collected with the BESIII spectrometer at center-of-mass energies between 4.600 GeV and 4.699~GeV. The amplitude analysis identifies four intermediate resonances $\Lambda(1830)^{0}$, $\Lambda(2000)^{0}$, $N(1440)^{+}$ and $N(1650)^{+}$. We measure the products of branching fractions:
\begin{eqnarray*}
\begin{aligned}
\mathcal{B}(\Lambda_{c}^{+} \rightarrow \Lambda(1830)^{0}\pip)\cdot \mathcal{B}(\Lambda(1830)^{0} \rightarrow n \Ks) = (2.4\pm0.7\pm0.5\pm0.1)\times 10^{-3},\\
\mathcal{B}(\Lambda_{c}^{+} \rightarrow \Lambda(2000)^{0}\pip)\cdot \mathcal{B}(\Lambda(2000)^{0} \rightarrow n \Ks)  = (5.4\pm1.7\pm1.1\pm0.3)\times 10^{-3},\\
\mathcal{B}(\Lambda_{c}^{+} \rightarrow N(1440)^{+}\Ks)\cdot \mathcal{B}(N(1440)^{+} \rightarrow n \pip)  = (4.0\pm2.0\pm1.5\pm0.2)\times 10^{-3},\\
\mathcal{B}(\Lambda_{c}^{+} \rightarrow N(1650)^{+}\Ks)\cdot \mathcal{B}(N(1650)^{+} \rightarrow n \pip)  = (5.2\pm1.1\pm0.8\pm0.3)\times 10^{-3},\\
\end{aligned}
\end{eqnarray*}
where the first uncertainties are statistical, the second are systematic, and the third are related to the normalizing branching fraction $\mathcal{B}(\sigmode)$. In addition, we measure the decay asymmetry parameters: 
\begin{eqnarray*}
\alpha_{\Lambda(1830)^{0}\pip} & = &  \phantom{-}0.93^{+0.07}_{-0.72}\pm0.07, \\
\alpha_{\Lambda(2000)^{0}\pip} & = & -0.91^{+0.56}_{-0.09}\pm0.14, \\
\alpha_{N(1440)^{+}\Ks} & = & -0.13\pm0.43\pm0.35, \\
\alpha_{N(1650)^{+}\Ks} & = & -0.42\pm0.45\pm0.29.
\end{eqnarray*}
}

\keywords{$\ee$ collision, Charmed baryon, Hadronic decay, Amplitude analysis}

\arxivnumber{XXXX.XXXXX}

\maketitle
\flushbottom

\section{Introduction}
\label{sec:introduction}
\hspace{1.5em} The MARKII experiment discovered the ground-state charmed baryon $\Lambda_{c}^{+}$ in 1979~\cite{Abrams:1979iu}. Its decay processes provide valuable information on the intrinsic dynamics underlying fundamental interactions~\cite{Li:2025nzx}. Although approximately 90\% of its decay channels have been observed~\cite{Gronau:2018vei}, intermediate resonant decay processes remain poorly characterized~\cite{Li:2025nzx}. This indicates that our current understanding of $\lcp$ decay dynamics is incomplete.

Lattice QCD calculations of $\lambdacp$ decays are highly challenging, primarily due to the complex non-perturbative effects inherent in the three-quark system involving charm quarks. Moreover, the factorization approach, which was successfully applied to bottom-hadron decays, does not adequately describe $\Lambda_{c}^{+}$ decays, as indicated by recent measurements of the $\lcp\to nK_S^0\pi^+$ decay~\cite{BESIII:2023pia}. Therefore, improved theoretical predictions require additional experimental input. In particular, measurements of the branching fractions of $\Lambda_{c}^{+}$ decays via intermediate resonances.

Despite the absence of numerical calculations, recent theoretical work~\cite{Li:2026umb} using the coupled-channel chiral unitary approach provides information about the characteristic intermediate resonances in the $\lcp\to nK_S^0\pi^+$ decay. This theoretical work shows a narrow peak from $N(1535)^+$ in the $n\pip$ invariant-mass spectrum and a distinct dip from $\Lambda(1670)^0$ in the $n\Ks$ invariant-mass spectrum. Therefore, to test these theoretical predictions, an amplitude analysis of this channel is essential to extract the contributions from intermediate resonances.

As pointed out in Ref.~\cite{Lu:2016ogy}, isospin symmetry imposes approximate constraints on the branching fractions (BFs) of the three isospin-related channels: $pK^{-}\pi^{+}$, $p\bar{K}^{0}\pi^{0}$ and $n\bar{K}^{0}\pi^{+}$. A more rigorous test of isospin symmetry becomes possible if the intermediate resonance contributions to these channels are well measured. These measurements allow precise tests of proportional relationships predicted by isospin among the BFs of the intermediate resonances. To date, the BFs of intermediate resonances in $\Lambda_{c}^{+} \to pK^{-}\pi^{+}$ have been reported by LHCb~\cite{LHCb:2022sck} and an amplitude analysis of $\Lambda_{c}^{+} \to pK_{S}^{0}\pi^{0}$ has been performed by BESIII~\cite{PWApkspi0}. As the last unmeasured channel among these three isospin-related modes, the present study measures the products of BFs for the intermediate resonances, enabling a comparative analysis to test isospin symmetry.

In this paper, we perform an amplitude analysis of charmed-baryon hadronic decay $\sigmode$. The analysis uses $e^+ e^-$ annihilation data corresponding to a total integrated luminosity of 4.5~$\rm{fb}^{-1}$ collected with the BESIII detector at center-of-mass~(c.m.) energies from 4.600~GeV to 4.699~GeV. Table~\ref{tab:lumi} lists the integrated luminosities of each c.m. energy point. We measure the fractions and decay asymmetries of four intermediate resonant processes through $\Lambda(1830)^{0}$, $\Lambda(2000)^{0}$, $N(1440)^{+}$ and $N(1650)^{+}$. Charge conjugation is implied throughout this paper.

\begin{table*}[!htbp]
\caption{Data sample c.m. energies and integrated luminosities~\cite{BESIII:2015qfd,BESIII:2022ulv}.}
\setlength{\abovecaptionskip}{1.2cm}
\setlength{\belowcaptionskip}{0.2cm}
\label{tab:lumi}
\begin{center}
\vspace{-0.0cm}
\begin{tabular}{c|c}
\hline\hline
$\sqrt{s}$ (GeV) & Luminosity ($\mathrm{pb}^{-1}$)\\
\hline
4.600 & $\phantom{0}586.9\pm0.1\pm3.9$\\
4.612 & $\phantom{0}103.8\pm0.1\pm0.6$\\
4.628 & $\phantom{0}521.5\pm0.1\pm2.8$\\
4.641 & $\phantom{0}552.4\pm0.1\pm2.9$\\
4.661 & $\phantom{0}529.6\pm0.1\pm2.8$\\
4.682 & $1669.3\pm0.2\pm8.8$\\
4.699 & $\phantom{0}536.4\pm0.1\pm2.8$\\
\hline\hline
\end{tabular}
\end{center}
\end{table*}
\vspace{-0.0cm}

\section{BESIII experiment and Monte Carlo simulation}
\label{sec:detector}
\hspace{1.5em} The $\uchyph=0$BESIII detector~\cite{BESIIIdetector} records symmetric $\ee$ collisions provided by the $\uchyph=0$BEPCII storage ring~\cite{BEPCII}. It has collected large data samples at c.m.\hspace{-0.2em}  energies ranging from 2.00 to 4.95~GeV~\cite{BESIII:2020nme}.  The cylindrical core of the $\uchyph=0$BESIII detector covers 93\% of the full solid angle and consists of a helium-based multilayer drift chamber (MDC), a plastic scintillator time-of-flight system (TOF) and a CsI(Tl) electromagnetic calorimeter (EMC), which are all enclosed in a superconducting solenoidal magnet providing a 1.0-T magnetic field. The solenoid is supported by an octagonal flux-return yoke with resistive-plate-counter muon-identification modules interleaved with steel. The momentum resolution of charged particles with a momentum of $1\,\gevc$ is $0.5\%$ and the resolution of the specific energy loss $\mathrm{d}E/\mathrm{d}x$ in the MDC is $6\%$ for electrons from Bhabha scattering. The EMC measures photon energies with a resolution of $2.5\%$ ($5\%$) at $1\,\gev$ in the barrel (end-cap) region.  The time resolution in the TOF barrel region is 68~ps, while that in the end-cap region was initially 110~ps. The end-cap TOF system was upgraded in 2015 using multi-gap resistive plate chamber technology, providing a time resolution of 60 ps, which benefits 87\% of the data used in this analysis~\cite{etof1,etof2,etof3}.

Simulated samples produced with {\sc geant4}-based~\cite{geant4} Monte Carlo (MC) software, which includes the geometric description of the BESIII detector and the detector response performance~\cite{GDMLMethod,BesGDML,Huang:2022wuo}, are used to determine detection efficiencies and estimate background contributions. The {\sc kkmc} generator~\cite{ref:kkmc1,ref:kkmc2} simulates beam-energy spread and initial-state radiation (ISR) in $e^+e^-$ annihilation.  The inclusive MC samples correspond to an integrated luminosity around forty times larger than that of the data samples. These samples include the production of $\lcp\lcm$ pairs, open-charm processes, the ISR production of vector charmonium(-like) states and the continuum processes incorporated in {\sc  kkmc}~\cite{ref:kkmc1,ref:kkmc2}.  The known decay modes are modelled with {\sc evtgen}~\cite{ref:evtgen1,ref:evtgen2} using BFs reported by the Particle Data Group (PDG)~\cite{ParticleDataGroup:2024cfk} and the remaining unknown charmonium decays are modelled with {\sc lundcharm}~\cite{ref:lundcharm1,ref:lundcharm2}. Final-state radiation from charged final-state particles is incorporated using {\sc photos}~\cite{photos}. MC samples of the signal process $\sigmode$ are produced with a uniform phase-space (PHSP) distribution.  At each c.m.-energy point, $2\times10^{6}$ events are generated, except at 4.682 GeV, where the sample consists of $4\times10^{6}$ events to reflect the larger integrated luminosity of that data set.

\section{Event selection}
\label{sec:selection}
\hspace{1.5em} The c.m. energies of the data sets are just above the production threshold for $\lcp\lcm$ pairs. Therefore, their production environment is clean because no accompanying particles are produced. Thus, a double-tag strategy is adopted~\cite{MARK-III:1985hbd}, which means that both $\lcp$ and $\lcm$ baryons are reconstructed, with the $\lcm$ candidate called the tag side and the $\lcp$ candidate called the signal side. Due to the detector's limited capability to reconstruct neutral hadrons~\cite{nwab181}, the signal-side neutron (antineutron) is not required to be detected directly; its momentum is calculated using four-momentum conservation.

\subsection{Tag-side $\lcm$ candidate selection}
\hspace{1.5em} The tag-side ($\lcm$) is reconstructed with eleven tag modes, $\lcm\to\Modea$, $\Modeb$, $\Modec$, $\Moded$, $\Modee$, $\Modeaa$, $\Modebb$, $\Modedd$, $\Modeaaa$, $\Modeccc$, and $\Modeddd$. The subsequent decay modes are $\Ks\to\pip\pim$, $\bar{\Sigma}^{0}\to\gamma\bar{\Lambda}$, $\bar{\Lambda}\to\bar{p}\pip$, $\bar{\Sigma}^{-}\to\bar{p}\pi^{0}$, and $\pi^{0}\to\gamma\gamma$. The tag-side candidates are reconstructed from the combinations of charged tracks and photon candidates recorded by the detector that satisfy the following selection criteria.

Charged-particle tracks detected in the MDC are required to have a polar angle $\theta$ in the range of $|\cos\theta|<0.93$, where $\theta$ is defined with respect to the $z$-axis, which is defined as the symmetry axis of the MDC. For tracks not originating from $\bar{\Lambda}$ or $\Ks$ decays, the distance of the closest approach to the interaction point (IP) is required to be less than 10~cm along the $z$-axis ($V_z$) and less than 1~cm in the perpendicular plane ($V_r$). Particle identification (PID) for charged particles combines information from specific-ionization energy loss ($\mathrm{d}E/\mathrm{d}x$) in the MDC  and the time-of-flight measured in the TOF into a likelihood value $\mathcal{L}(h)$ for each hadron $h$ hypothesis, where $h=p$, $K$, or $\pi$. Tracks are identified as protons if the proton hypothesis has the greatest likelihood ($\mathcal{L}(p)>\mathcal{L}(K)$, $\mathcal{L}(p)>\mathcal{L}(\pi)$ and $\mathcal{L}(p)>0$), as pions if $\mathcal{L}(\pi)>\mathcal{L}(K)$ and $\mathcal{L}(\pi)>0$, and as kaons if $\mathcal{L}(K)>\mathcal{L}(\pi)$ and $\mathcal{L}(K)>0$. 

Photon candidates are reconstructed using isolated electromagnetic showers in the EMC. The energy of each shower must be greater than $25\,\mev$ in the barrel region ($|\cos \theta|< 0.80$) and greater than $50\,\mev$ in the end-cap region ($0.86 <|\cos \theta|< 0.92$). To reject fake photons arising from electronic noise, beam background and showers unrelated to the event, the difference between the EMC time and the event start time~\cite{Guan:2013jua} must be within 700 ns. For $\piz$ candidates, the invariant mass of the photon pair is required to be within $0.115<M_{\gamma\gamma}<0.150\,\gevcc$. To improve the momentum resolution, we perform a one-constraint (1C) kinematic fit that constrains the invariant mass of the photon pair to the known $\piz$ mass~\cite{ParticleDataGroup:2022pth}. The updated $\piz$ momentum is used in the subsequent analysis.

The $\bar{\Lambda}$ ($\Ks$) candidates are reconstructed with two oppositely charged particles identified as $\bar{p}$ ($\pi^-$) and $\pi^+$. The tracks must satisfy $|V_z| < 20$~cm, but no $V_r$ requirement is imposed. For the proton candidate, we apply the previously described PID requirement, while no PID requirement is imposed on the pion candidate. A vertex fit constrains $\bar{p}\pip$ pairs to originate from a common vertex; we select candidates with a fit $\chi^2$ less than 100. An additional fit is performed by constraining the momentum of the $\bar{p}\pip$ pair to align with the direction from the IP to the $\bar{\Lambda}$-decay vertex; the fitted decay length ($L$) must be greater than twice its uncertainty ($\sigma_{L}$). Furthermore, the $\bar{\Lambda}~(\Ks)$-candidate's invariant mass must be within $1.111<M_{\bar{p}\pip}<1.121\,\gevcc$ ($0.487<M_{\pip\pim}<0.511\,\gevcc$). The invariant mass of a $\bar{\Sigma}^0\to\gamma\bar{\Lambda}$ [$\bar{\Sigma}^-\to\bar{p}\pi^{0}$] candidate must lie within $(1.179,\,1.203)\,\gevcc$ [$(1.176,\,1.200)\,\gevcc$].

To identify tag-side candidates, we use the beam-constrained mass $$\mbc \equiv \sqrt{{\ebeam}^2/c^4-\left|\mathbf{p}\right|^2/c^2}\;$$ and the beam-energy difference $$\dE \equiv E-\ebeam\;,$$ where $\ebeam$ is the beam energy, $\mathbf{p}$ and $E$ are the reconstructed c.m. momentum and energy of the tag-side candidate. An $\lcm$ candidate must satisfy a tag-mode-dependent $\dE$ requirement, which is determined by the smallest $\dE$ interval that contains 97\% of signal MC events. If more than one candidate satisfies the above requirements for a specific tag mode, the one with the minimum $|\dE|$ is kept. For all tag modes, $\mbc$ must be within $(2.280,\,2.296)\,\gevcc$.

For the $\lcm \to \bar{p}\Ks\piz$ tag mode, candidate events with $M_{\bar{p}\pip} \in (1.110,\,1.120)\,\gevcc$ and $M_{\bar{p}\piz} \in (1.170,\,1.200)\,\gevcc$ are vetoed to avoid double counting with the $\lcm\to \bar{\Lambda}\pim\piz$ or $\lcm\to \bar{\Sigma}^-\pip\pim$ tag mode, respectively. For the $\lcm\to \bar{\Sigma}^-\pip\pim$ tag mode, we veto candidates with $M_{\pip\pim} \in (0.480,\,0.520)\,\gevcc$ and $M_{\bar{p}\pip} \in (1.110,\,1.120)\,\gevcc$. Finally, when selecting $\lcm\to \bar{p}\Ks\pip\pim$, $\lcm\to \bar{\Lambda}\pim\pip\pim$ and $\lcm\to \bar{\Sigma}^-\piz$ candidates, we reject those with $M_{\bar{p}\pip} \in (1.110,\,1.120)\,\gevcc$, $M_{\pip\pim} \in (0.480,\,0.520)\,\gevcc$ and $M_{\piz\piz} \in (0.480,\,0.520)\,\gevcc$, respectively.

\subsection{Signal-side $\lcp$ candidate selection}
\hspace{1.5em}
We reconstruct signal-side $\sigmode$ candidates with tracks not used in the reconstruction of tag-side $\lcm$ candidates. The criteria for signal-side $\Ks(\to \pip\pim)$ reconstruction are identical to those of tag-side $\Ks$ candidates. If there are multiple combinations of $\pip\pim$, we select the one with the largest $L/\sigma_{L}$. For the pion directly from the $\lcp$ decays, we apply identical track and PID criteria as used in the tag-side reconstruction. Except for these three pions ($\Ks\to \pip\pim$ and $\pip$), we require that no other tracks are reconstructed in the detector.

To extract signal yields, a kinematic variable $M_{\mathrm{miss}}^{2}$ is calculated from the $\pip$ and $\Ks$ candidates:
\begin{equation}
\label{eq:Mmiss.1}
M_{\rm miss}^{2} = E_{\rm miss}^{2}-c^{2}|\mathbf{p}_{\rm miss}|^{2},
\end{equation}
where $E_{\rm miss}$ and $\mathbf{p}_{\rm{miss}}$ are the missing energy and momentum carried away by the undetected neutron. Here, we define
\begin{equation}
	\label{eq:Mmiss.2}
	E_{\rm miss} = E_{\rm beam}-E_{K_{S}^{0}}-E_{\pi^{+}},
\end{equation}
where $E_{K_{S}^{0}}$ and $E_{\pi^{+}}$ are measured energies of $K_{S}^0$ and $\pi^{+}$ in the $e^+e^-$ c.m. system, respectively, and we define
\begin{equation}
	\label{eq:Mmiss.3}
	\mathbf{p}_{\rm miss} = \mathbf{p}_{\Lambda_{c}^{+}}-\mathbf{p}_{K_{S}^{0}}-\mathbf{p}_{\pi^{+}},
\end{equation}
where $\mathbf{p}_{\Lambda_{c}^{+}}$, $\mathbf{p}_{K_{S}^{0}}$ and $\mathbf{p}_{\pi^{+}}$ are momenta of $\Lambda_{c}^{+}$, $K_{S}^{0}$, $\pi^{+}$ in the $e^+e^-$ c.m. system, respectively. The definition of $\mathbf{p}_{\Lambda_{c}^{+}}$ is
\begin{equation}
	\label{eq:Mmiss.4}
	\mathbf{p}_{\Lambda_{c}^{+}}=-\hat{\mathbf{p}}_{\rm tag}\sqrt{E_{\rm beam}^{2}/c^2-m_{\Lambda_{c}^{+}}^{2}c^2},
\end{equation}
where $\hat{\mathbf{p}}_{\rm tag}$ is the direction of the tag-side $\lambdacm$ candidate and $m_{\Lambda_{c}^{+}}$ is the known mass of $\Lambda_{c}^{+}$~\cite{ParticleDataGroup:2024cfk}. The $M_{\rm{miss}}^{2}$ distribution of signal candidates peaks at $0.8828~{\rm{GeV}^2}/c^4$, which is the known squared mass of the neutron~\cite{ParticleDataGroup:2024cfk}.

Amplitude analyses require data samples without peaking backgrounds and minimal combinatorial background. Therefore, to suppress peaking backgrounds within the signal region of the $M_{\rm miss}^{2}$ spectrum, we veto events with the $\lambdacp\to\Sigmap\pip\pim$, $\Sigmap\Ks$, and $\Sigma^{-}\pip\pip$ decay modes. To veto peaking backgrounds involving $\Sigma^+$ or $\Sigma^-$, we reject events if either of the two $\pi^+$ candidates satisfy $M_{n\pi^+}-M_n\in(0.235,\,0.265)\,\gevcc$, or if the $\pi^-$ candidate satisfies $M_{n\pi^-}-M_n\in(0.240,\,0.270)\,\gevcc$.
The calculated value of $\mathbf{p}_{\rm miss}$, together with the known neutron mass $M_n$~\cite{ParticleDataGroup:2024cfk}, is used to reconstruct the neutron four-momentum, from which we determine the invariant mass $M_{n\pi}$.

\begin{figure}[htbp]
\setlength{\abovecaptionskip}{-1pt}
\setlength{\belowcaptionskip}{10pt}
\centering
\includegraphics[trim = 9mm 0mm 0mm 0mm, width=0.31\textwidth]{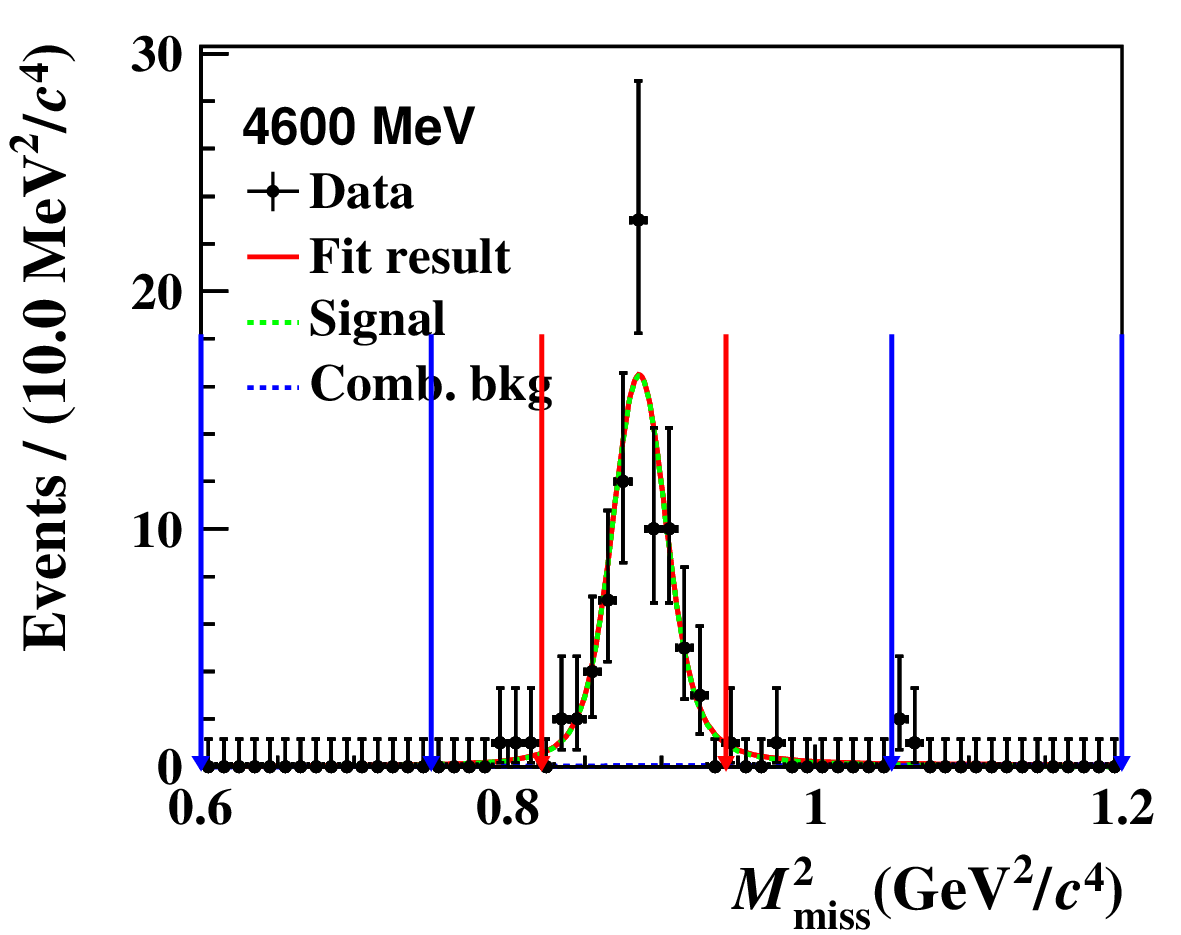}
\includegraphics[trim = 9mm 0mm 0mm 0mm, width=0.31\textwidth]{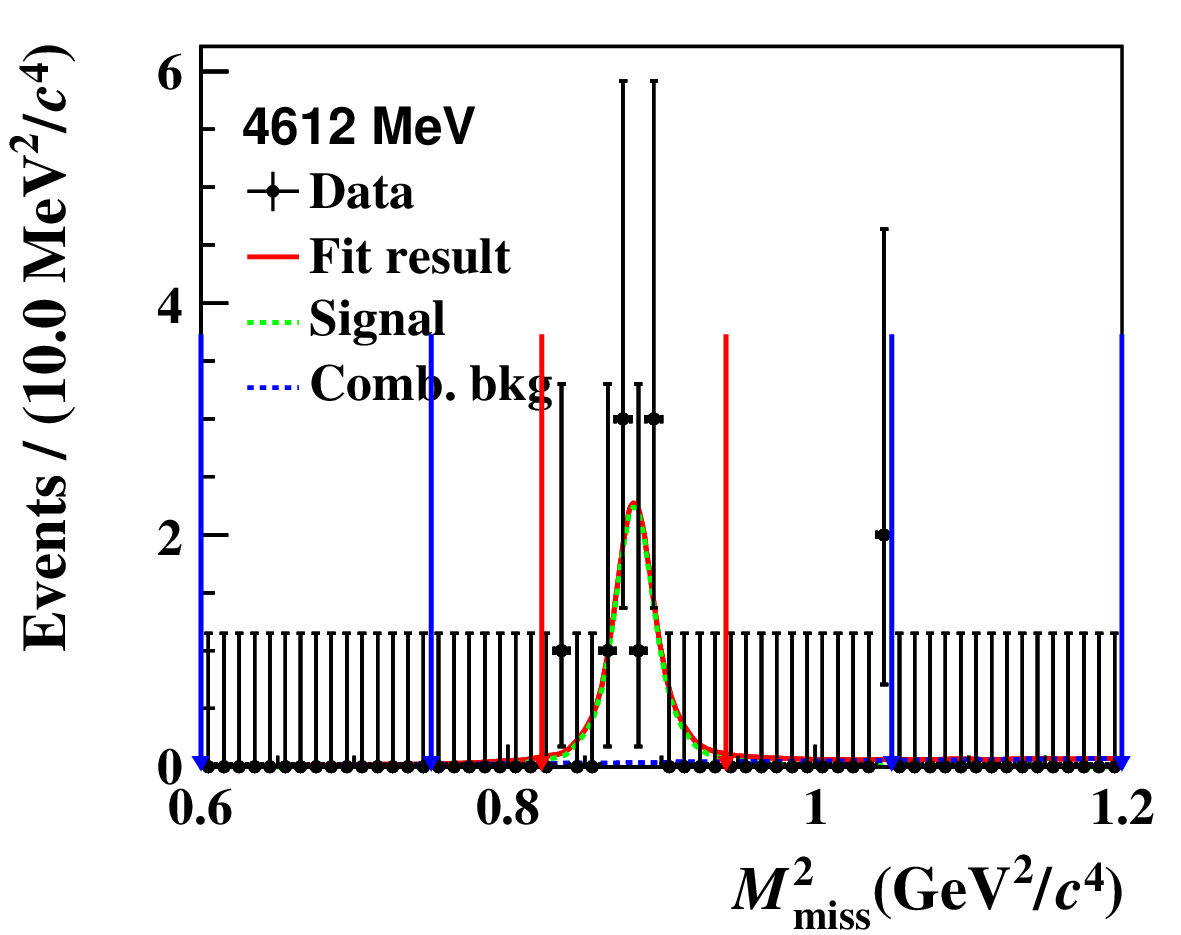}
\includegraphics[trim = 9mm 0mm 0mm 0mm, width=0.31\textwidth]{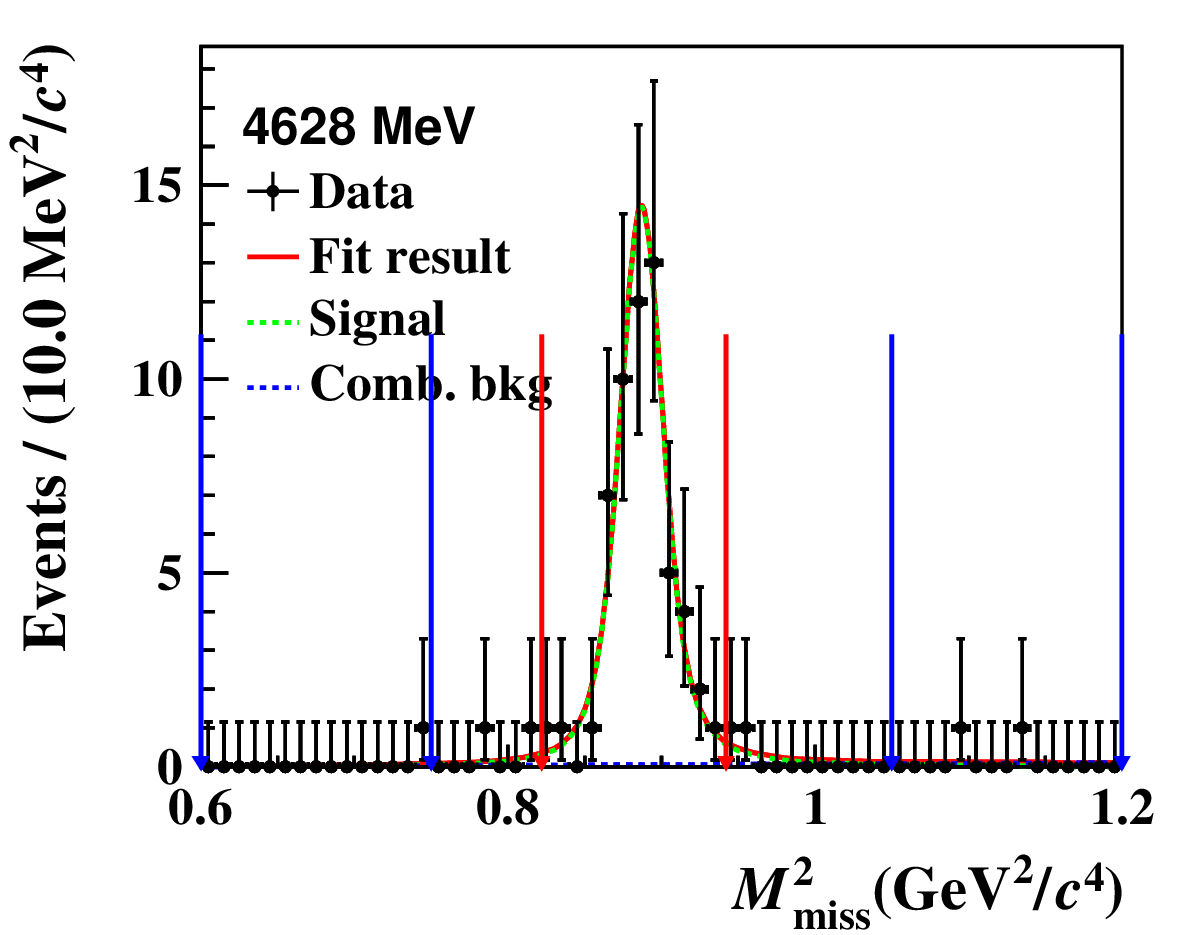}\\
\includegraphics[trim = 9mm 0mm 0mm 0mm, width=0.31\textwidth]{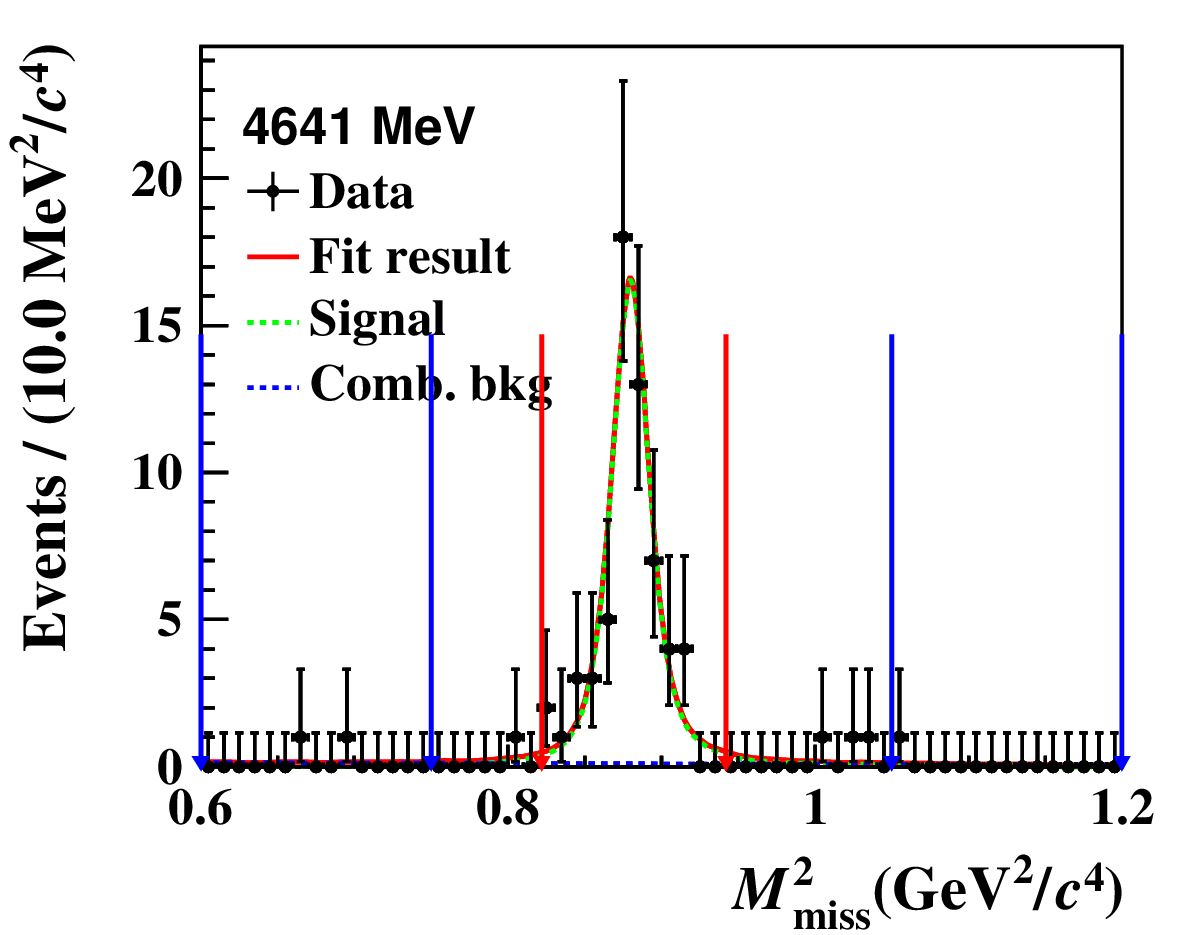}
\includegraphics[trim = 9mm 0mm 0mm 0mm, width=0.31\textwidth]{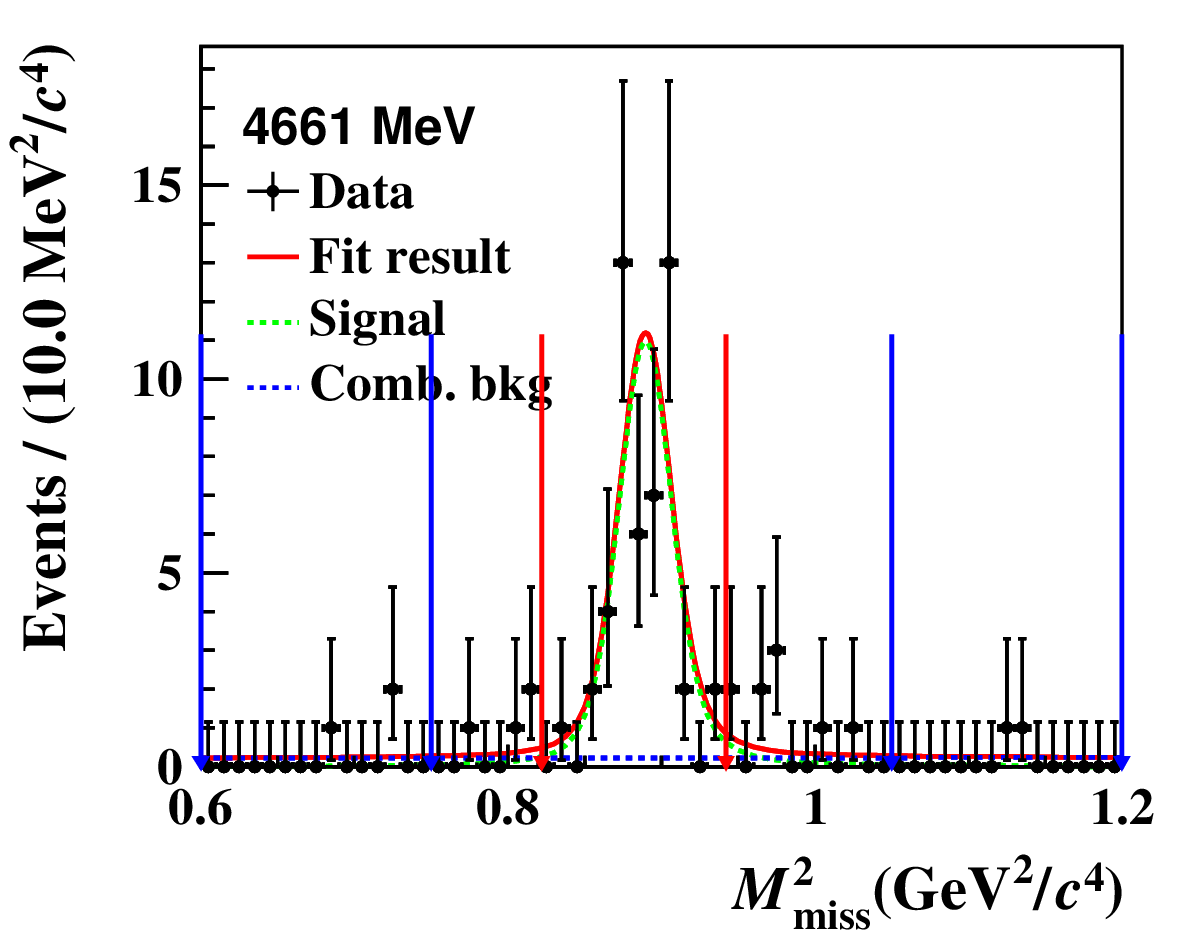}
\includegraphics[trim = 9mm 0mm 0mm 0mm, width=0.31\textwidth]{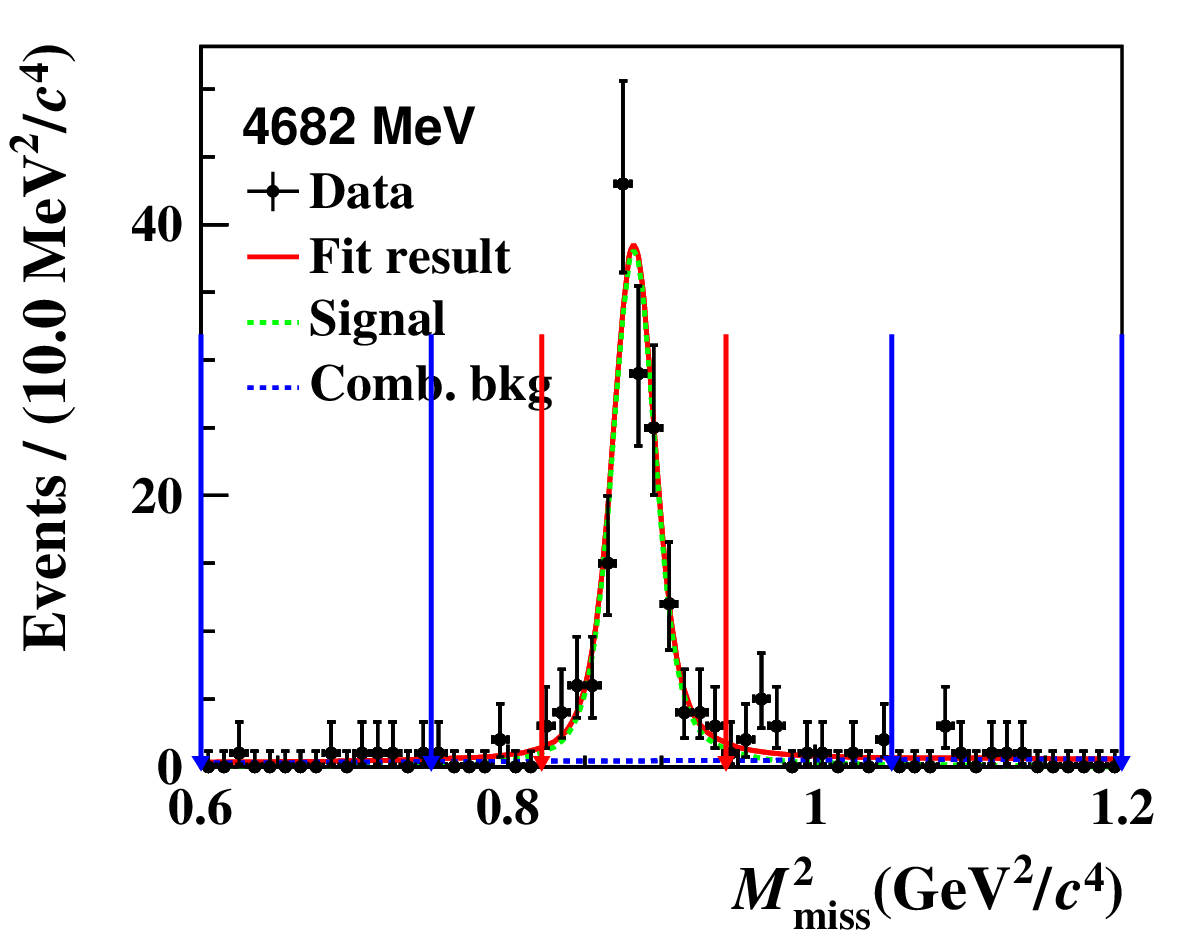}\\
\includegraphics[trim = 9mm 0mm 0mm 0mm, width=0.31\textwidth]{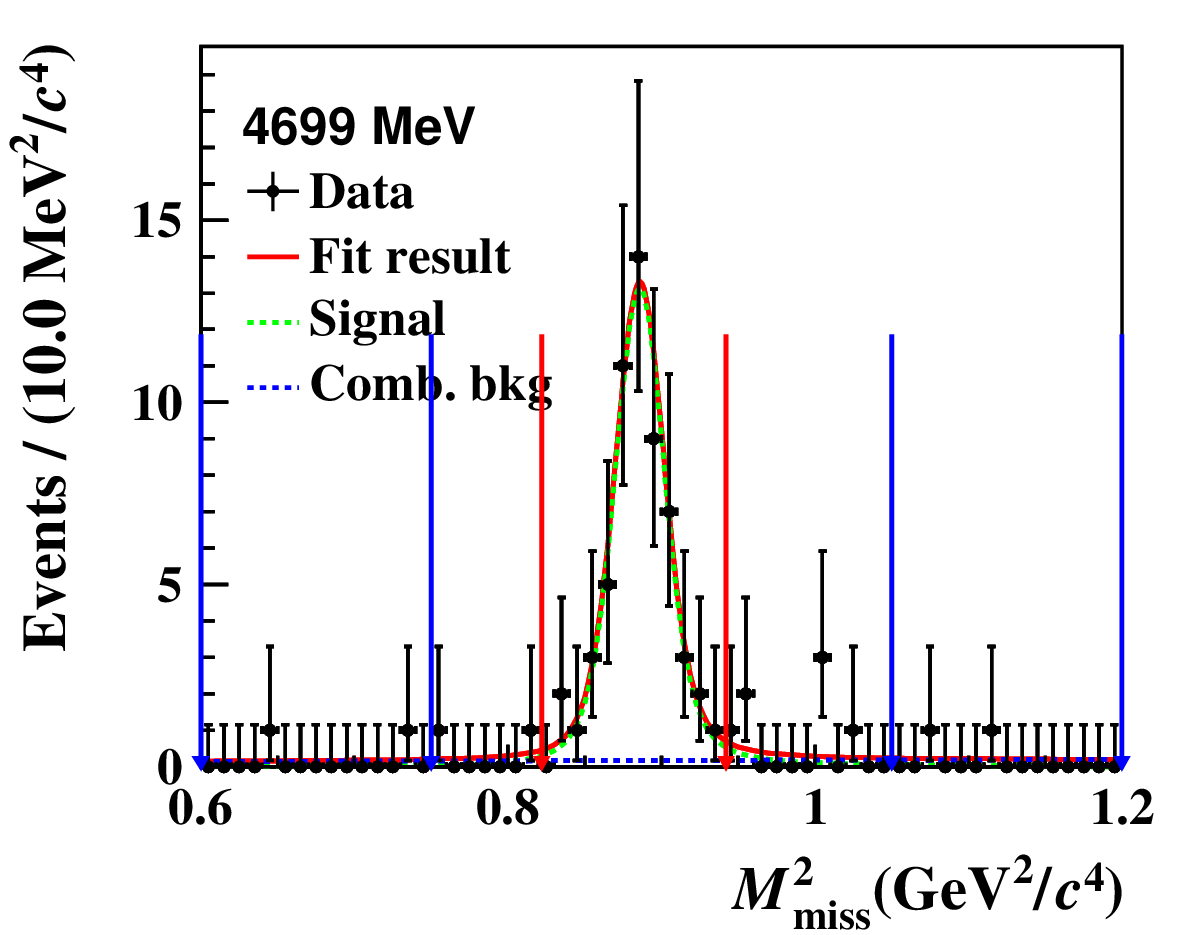}
\caption{$M_{\rm miss}^2$ distribution fit results for each energy point. The regions between the two red vertical lines indicate the $M_{\rm miss}^2$ signal region, and the regions between four blue vertical lines indicate the upper and lower $M_{\rm miss}^2$ sideband regions.}
\label{fig:fitmiss2}
\end{figure}

To determine the signal yield at each energy point, we perform a one-dimensional extended maximum likelihood fit to the $M_{\rm miss}^{2}$ distribution. The signal probability density function (PDF) is derived from the PHSP signal MC sample generated at the corresponding energy point; this PDF is convolved with a Gaussian function with free parameters to describe any resolution discrepancy between data and MC samples. The Gaussian parameters are different at each energy point. The background PDF is described by a straight line. Figure~\ref{fig:fitmiss2} shows the fit results and Table~\ref{tab:fitmiss2} lists the numerical results. Based on these results, we determine that the $M_{\rm miss}^{2}$ signal region is $(0.822,\,0.942)\,{\rm{GeV}^2}/c^4$ and the sideband regions are $(0.60,\,0.75)$ and $(1.05,\,1.20)\,{\rm{GeV}^2}/c^4$.

\vspace{-0.0cm}

\begin{table*}[!htbp]
\caption{Signal and background yields in the $M_{\rm miss}^2$ signal region at each energy points, along with signal purities and background fractions.}
\setlength{\abovecaptionskip}{1.2cm}
\setlength{\belowcaptionskip}{0.2cm}
\label{tab:fitmiss2}
\begin{center}
\vspace{-0.0cm}
\begin{tabular}{c|ccccc}
	\hline \hline
	$\sqrt{s}$ (GeV) & Signal yield & Background yield & Purity (\%) & Background fraction (\%) \\
	\hline
	4.600  & $79\pm11$ & $0.3\pm0.5$ & $99.6\pm0.6$ & $0.4\pm0.6$\\
	4.612  & $8\pm3$  & $0.4\pm0.3$  & $95.0\pm4.0$ & $5.0\pm4.0$\\
	4.628  & $57\pm7$ & $0.6\pm0.5$ & $99.0\pm0.9$ & $1.0\pm0.9$\\
	4.641  & $58\pm8$ & $1.1\pm0.7$ & $98.1\pm1.2$ & $1.9\pm1.2$\\
	4.661  & $51\pm8$ & $2.7\pm1.0$ & $94.9\pm1.9$ & $5.1\pm1.9$\\
	4.682  & $150\pm13$ & $5.1\pm1.4$ & $96.7\pm0.9$ & $3.3\pm0.9$\\
	4.699  & $58\pm8$  & $2.0\pm0.9$ & $96.6\pm1.5$ & $3.4\pm1.5$\\
	\hline\hline
\end{tabular}
\end{center}
\end{table*}
\vspace{-0.0cm}

\section{Amplitude analysis}
\label{sec:PWA}
\hspace{1.5em} 

After applying all the criteria mentioned above, we select $462\pm58$ signal events to be used for the amplitude analysis. To improve the momentum resolution, we perform an additional 4C kinematic fit, under the hypothesis of $\ee \to \lcp (\to n\Ks\pi^+) \lcm$. The fit constrains the $\pip\pim$ invariant mass to the known $\Ks$ mass, $M_{\mathrm{miss}}$ to the neutron mass, and the invariant masses of $n\Ks\pip$ and tag-side $\lcm$ to the $\Lambda_{c}^{+}$ mass. 

The decay amplitude is constructed using the helicity amplitude formalism; the full procedure is implemented based on the open-source TF-PWA framework~\cite{Jiang:tfpwa}. The amplitude is defined in the $e^+e^-$ rest frame to which the four-momenta of all final-state particles are boosted. Assuming $C\!P$ conservation, the parameters of the $\lcm$-decay amplitude are related to those of the $\lcp$-decay amplitude by a $C\!P$ transformation of $\lcm$ candidates.

\subsection{Helicity angle definitions}
\label{sec:angle}

\begin{figure}[H]
\setlength{\abovecaptionskip}{-1pt}
\setlength{\belowcaptionskip}{10pt}
\centering
\includegraphics[trim = 9mm 0mm 0mm 0mm, width=0.6\textwidth]{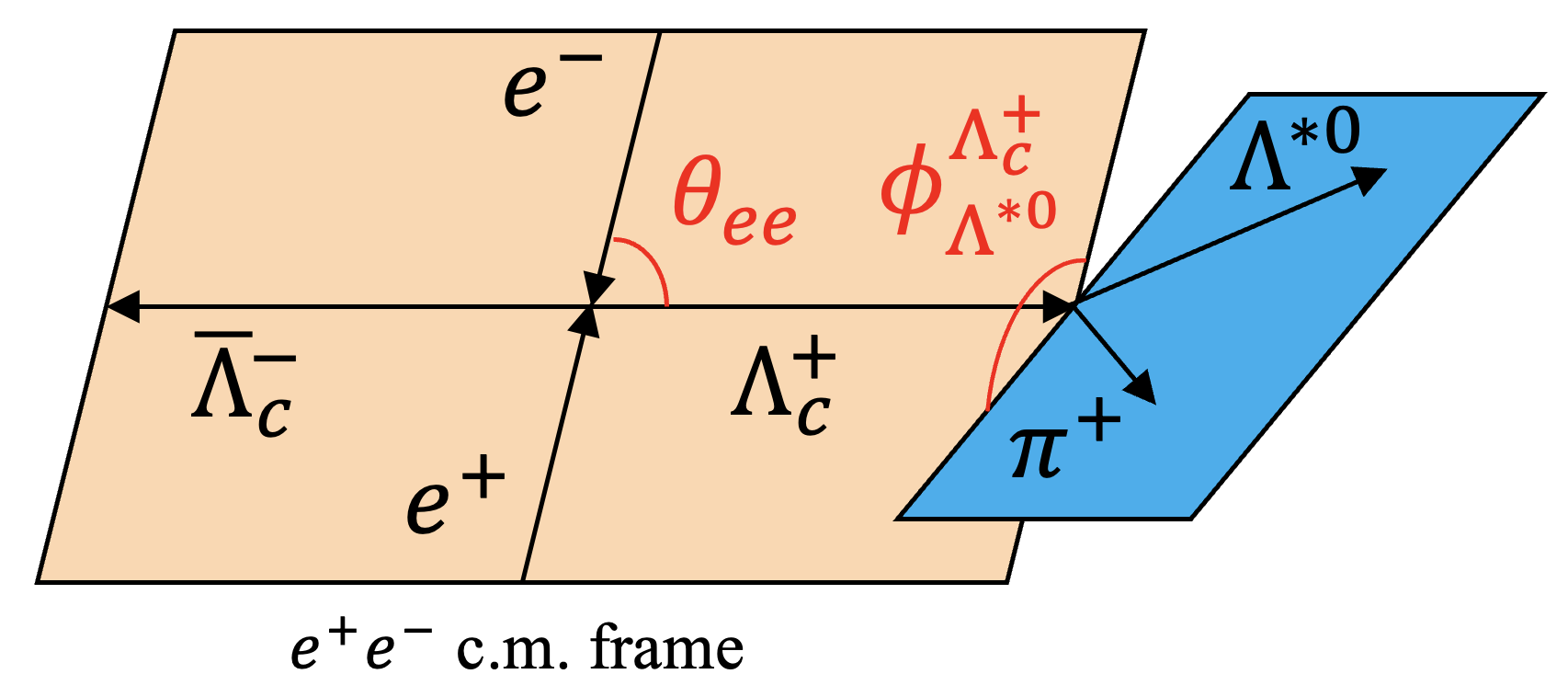}\\
\vspace{0.6cm}
\includegraphics[trim = 9mm 0mm 0mm 0mm, width=0.6\textwidth]{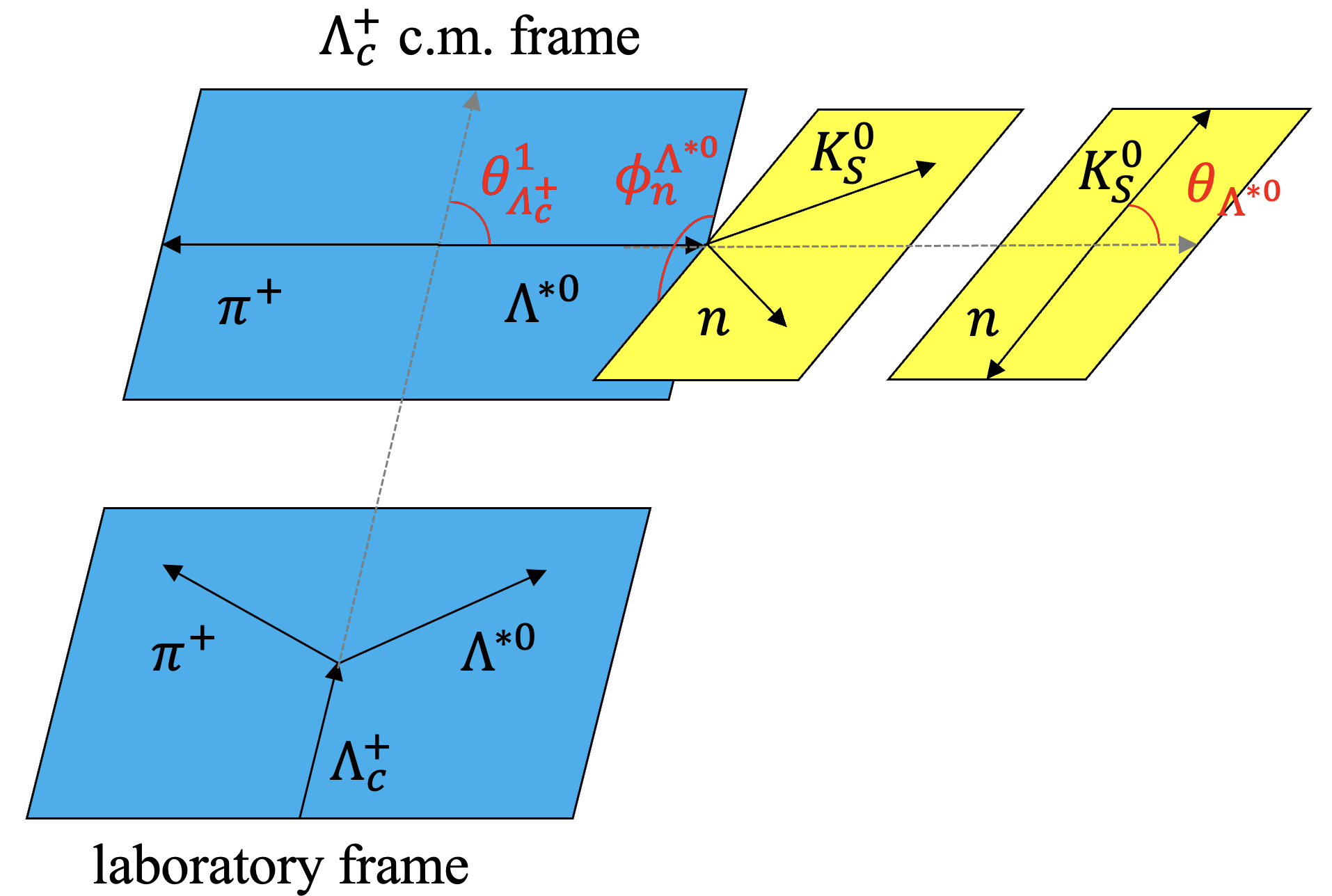}\\
\vspace{0.6cm}
\includegraphics[trim = 9mm 0mm 0mm 0mm, width=0.6\textwidth]{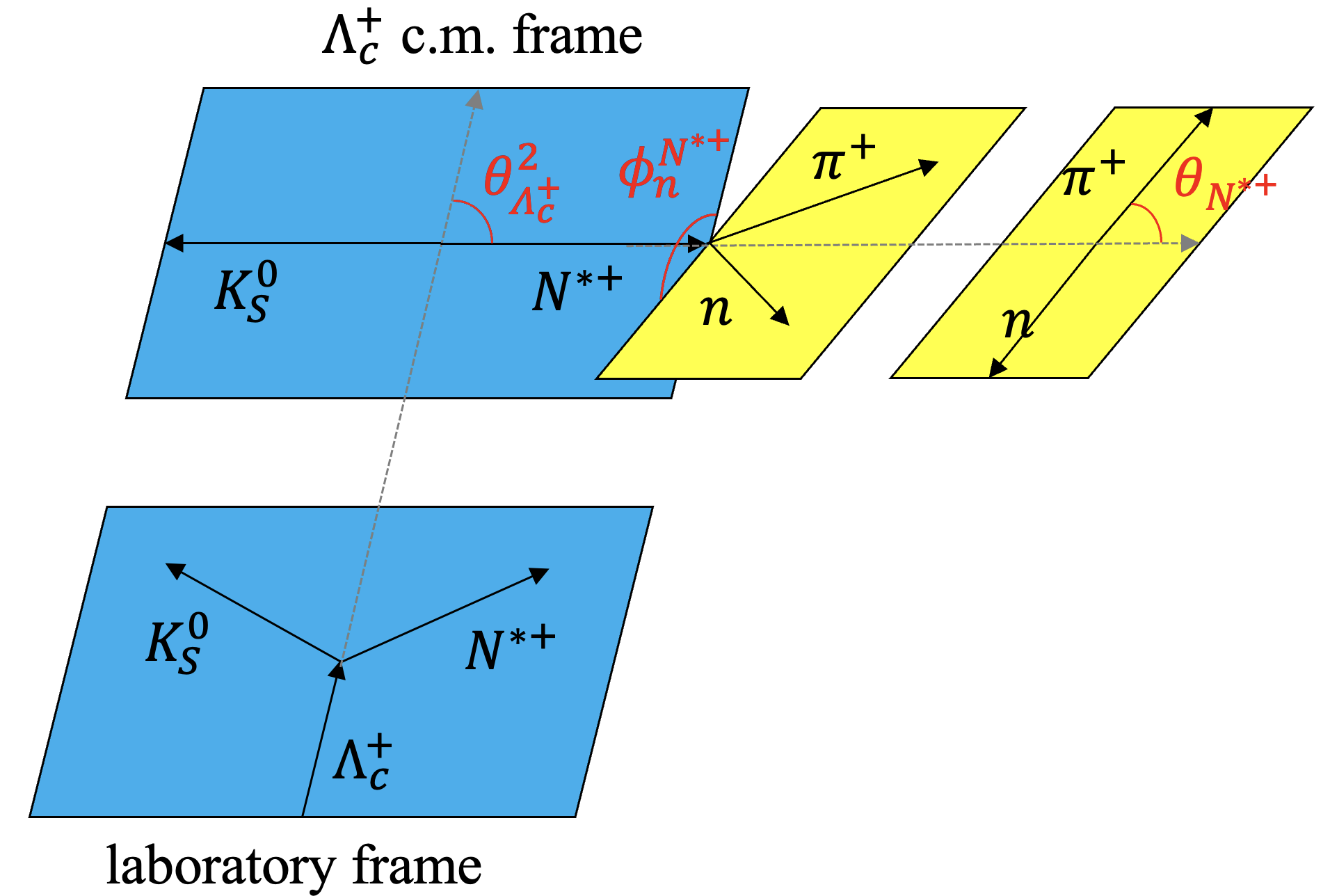}
\vspace{1.0\baselineskip}
\caption{Definitions of the helicity angles with different decay chains in the decay $\sigmode$. The notation $\theta_R$ denotes the $\theta$ helicity angle of the decay of resonance $R$. For the $\lcp$ helicity angle, the superscripts 1 and 2 are used to differentiate two decay chains. The notation $\phi_B^A$ denotes the $\phi$ helicity angle between the plane of $B$ and the plane of its mother particle $A$. By convention, the right-handed frame is chosen as the normal direction for a $\phi$ rotation, and the anti-direction rotation is denoted as $-\phi$ in the figure. The exact calculations of different helicity angles are implemented according to the conventions in Ref.~\cite{helicity}.}
\label{fig:angle}
\end{figure}

Decays involving resonances in the $K^0\pip$ system are doubly Cabibbo suppressed, and their contributions are neglected given the current statistics. The amplitude model for $\sigmode$ therefore includes the following two decay chains:
\begin{itemize}
\item $e^+e^-\to\lambdacp\lambdacm(\theta_{ee}),\lambdacp\to N^{*+}\Ks(\theta_{\lambdacp}^{1},\phi_{N^{*+}}^{\lambdacp}), N^{*+}\to n\pip(\theta_{N^{*+}}, \phi_{n}^{N^{*+}})$, 
\item $e^+e^-\to\lambdacp\lambdacm(\theta_{ee}),\lambdacp\to\Lambda^{*0}\pip(\theta_{\lambdacp}^{2},\phi_{\Lambda^{*0}}^{\lambdacp}), \Lambda^{*0}\to n \Ks(\theta_{\Lambda^{*0}}, \phi_{n}^{\Lambda^{*0}})$.
\end{itemize}
The corresponding helicity angle definitions are shown in Fig.~\ref{fig:angle}.
\vspace{-0.0cm}

\subsection{Helicity amplitude}
\label{sec:helicity}
\hspace{1.5em} To construct the decay amplitude of $\sigmode$, we use the helicity formalism~\cite{Chung:1971ri,Richman:1984gh} based on the isobar model~\cite{PhysRev.145.1229}, in which the three-body decay is described by two-step sequential quasi-two-body decays. For each two-body decay $0\to1+2$, the helicity amplitude is
\begin{eqnarray}
    A^{0\rightarrow 1+2}_{\lambda_0, \lambda_1, \lambda_2} = H_{\lambda_1, \lambda_2}^{0\rightarrow 1+2} D^{J_0 *}_{\lambda_0,\lambda_1-\lambda_2}(\phi, \theta, 0),
\label{eq:helicty1}
\end{eqnarray}
where the amplitude $H_{\lambda_1, \lambda_2}^{0\rightarrow 1+2}$ is given by the LS-coupling formula~\cite{Chung:1993da,Chung:1997jn,Chung:2003fk} along with barrier-factor terms
\begin{eqnarray}
    H_{\lambda_1, \lambda_2}^{0\rightarrow 1+2} = \sum_{l s} g_{ls} \sqrt{\frac{2l+1}{2J_0 + 1}} 
    \langle l 0s \delta| J_0\delta \rangle
    \langle J_1 \lambda_1 J_2\ -\lambda_2| s\delta \rangle \left(\frac{q}{q_0}\right)^l B_l'(q,q_0,d),
\label{eq:helicty2}
\end{eqnarray}
where $g_{ls}$ is the partial wave amplitude, $J_{i}$ and $\lambda_{i}$ are the spin and helicity of the particle $i$, respectively, and $\delta=\lambda_1-\lambda_2$ is the helicity difference.  Here, $q$ is the magnitude of the three-momentum of particle $1$ in the rest frame of particle $0$, which is calculated as
\begin{eqnarray}
    q = \frac{\sqrt{[m^2 - (m_1 +m_2)^2][m^2 - (m_1 - m_2)^2]}}{2 m}.
\label{eq:helicty3}
\end{eqnarray}
Here $m$, $m_1$ and $m_2$ are the masses of particles 0, 1 and 2, respectively. The normalization factor $q_0$ is calculated at the resonance pole mass.  The factor $B_l'(q,q_0,d)$ is the reduced Blatt-Weisskopf barrier factor~\cite{Blatt:1952ije,VonHippel:1972fg}, which is explicitly expressed as
\begin{eqnarray}
\begin{aligned}
    B_0'(q,q_0,d) & =1,\\
    B_1'(q,q_0,d) & =\sqrt{\frac{1+(q_0d)^2}{1+(qd)^2}},\\
    B_2'(q,q_0,d) & =\sqrt{\frac{9+3(q_0d)^2+(q_0d)^4}{9+3(qd)^2+(qd)^4}},\\
    B_3'(q,q_0,d) & =\sqrt{\frac{225+45(q_0d)^2+6(q_0d)^4+(q_0d)^6}{225+45(qd)^2+6(qd)^4+(qd)^6}}.
\end{aligned}
\label{eq:barrier}
\end{eqnarray}
In the Wigner $D$-function, $D^{J_0*}_{\lambda_0,\lambda_1-\lambda_2}(\phi, \theta, 0)$, $\phi$ and $\theta$ are helicity angles as shown in Fig.~\ref{fig:angle}. The definitions can be found in Ref.~\cite{Wang:2020giv}. In Eq.~\eqref{eq:barrier}, the radius of the centrifugal barrier $d$ is chosen as $d=1/q_{r}=3.69\,(\mathrm{GeV}/c)^{-1}$ corresponding to the hadron ``scale'' parameter $q_{r}=0.2708\,\gevc$, which is equivalent to the hadron radius $R=0.73\;\mathrm{fm}$ with a relation $q_{r}=(0.197321/R[\mathrm{fm}])\,\gevc$ ~\cite{Zou:2002ar}. For the decay chain including non-resonant ($N\!R$) intermediate state, we neglect barrier-factor and momentum terms in the LS-coupling formulation, so it is written as
\begin{eqnarray}
    H_{\lambda_1, \lambda_2}^{0\rightarrow 1+2} = \sum_{l s} g_{ls} \sqrt{\frac{2l+1}{2J_0 + 1}} 
    \langle l 0s \delta| J_0\delta \rangle
    \langle J_1 \lambda_1 J_2\ -\lambda_2| s\delta \rangle.
\label{eq:helicty_NR}
\end{eqnarray}

The amplitude for a $\lcp$ decay chain is constructed as the product of each two-body decay amplitude and its resonant propagator $R$. For example, in the sequential decay $\Lambda_c^{+} \rightarrow \Lambda^{*0}
\pi^+$, $\Lambda^{*0} \to n  \Ks$, the amplitude is written as
\begin{eqnarray}
    A_{\lambda_{\Lambda_c^{+}}, \lambda_n}^{\Lambda^{*0}}=\sum_{\lambda_{\Lambda^{*0}}} A_{\lambda_{\Lambda_c^{+}}, 0, \lambda_{\Lambda^{*0}}}^{\Lambda_{c}^{+} \rightarrow \Lambda^{*0}\pip} R_{\Lambda^{*0}}\left(M_{n\Ks}\right) A_{\lambda_{\Lambda^{*0}}, \lambda_n, 0}^{\Lambda^{*0} \rightarrow n\Ks}.
\label{eq:helicty4}
\end{eqnarray}
Similarly, for the decay via the $N^{*+}$ intermediate states, the amplitude is
\begin{eqnarray}
    A^{N^{*+}}_{\lambda_{\lcp},\lambda_n} = \sum_{\lambda_{N^{*+}}}A_{\lambda_{\Lambda_c^{+}}, \lambda_{N^{*+}}, 0}^{\Lambda_{c}^{+} \rightarrow N^{*+}\Ks} R_{N^{*+}}(M_{n\pip})A_{\lambda_{N^{*+}}, \lambda_{n},0}^{N^{*+} \rightarrow n \pi^{+}}.
\label{eq:helicty4b}
\end{eqnarray}
The propagator $R$ models the resonance.
For $N^{*+}$ and $\Lambda^{*0}$ resonances, we use a relativistic Breit-Wigner parametrization: 
\begin{eqnarray}
    R_{N^{*+}/\Lambda^{*0}}(m) = \frac{1}{m_0^2-m^2 -i m_0 \Gamma(m)},
\label{eq:helicty6}
\end{eqnarray}
where the mass-dependent width is
\begin{eqnarray}
    \Gamma(m) = \Gamma_0 \left(\frac{q}{q_0}\right)^{2l+1}\frac{m_0}{m} B_l'^2(q,q_0,d).
\label{eq:helicty7}
\end{eqnarray}
The amplitude of $\sigmode$ is the coherent sum of all possible resonance amplitudes: 
\begin{eqnarray}
\begin{aligned}
    \mathcal{A}_{\lambda_{\Lambda_c^{+}}, \lambda_n} & =\sum_{\lambda_n^{\prime}} A_{\lambda_{\Lambda_c^{+}}, \lambda_n^{\prime}}^{N^{*+}} D_{\lambda_n^{\prime}, \lambda_n}^{1 / 2}\left(\alpha_n, \beta_n, \gamma_n\right) \\
    & +\sum_{\lambda_n^{\prime}} A_{\lambda_{\Lambda_c^{+}}, \lambda_n^{\prime}}^{\Lambda^{*0}} D_{\lambda_n^{\prime}, \lambda_n}^{1 / 2}\left(\alpha_n^{\prime}, \beta_n^{\prime}, \gamma_n^{\prime}\right),
\end{aligned}
\label{eq:helicty13}
\end{eqnarray}
where the extra alignment $D$-functions are added to align the helicities of the final-state neutrons. Details of the alignment-angle calculations are in Appendix~\ref{app:align_angle} and in Ref.~\cite{Wang:2020giv}. 

When we consider the polarization of $\lcp$ in the process of $e^+e^-\to \gamma^*\to\lcp\lcm$ using the most general spin-$1/2$ density matrix, the total amplitude can be written as 
\begin{equation}
	\begin{aligned}
	|\mathcal{A}|^2 &= \sum_{\lambda_{\lambdacp},\lambda_{\lambdacp}^{\prime},\lambda_{n}} \rho_{\lambda_{\lambdacp},\lambda_{\lambdacp}^{\prime}} \mathcal{A}_{\lambda_{\lambdacp},\lambda_{n}} \mathcal{A}^{*}_{\lambda_{\lambdacp}^{\prime},\lambda_{n}} \\
	&= \sum_{\lambda_{n}} \left[
	\left(\begin{array}{cc}
		\mathcal{A}_{-1/2,\lambda_{n}} & \mathcal{A}_{1/2,\lambda_{n}}
	\end{array}\right)
	\cdot \frac{1}{2} \cdot
	\left(\begin{array}{cc}
		1+P_z & P_x-i P_y \\
		P_x+i P_y & 1-P_z
	\end{array}\right)
	\cdot
	\left(\begin{array}{c}
		\mathcal{A}^{*}_{-1/2,\lambda_{n}} \\
		\mathcal{A}^{*}_{1/2,\lambda_{n}}
	\end{array}\right)
	\right], 
	\end{aligned}
\end{equation}
where $P_x$, $P_y$ and $P_z$ are the polarization vector's components. According to Ref.~\cite{Chen:2019hqi}, the polarization components are
\begin{equation}
	P_z = P_x = 0~\mathrm{and}~P_y(\theta_{ee},\alpha_0,\Delta_0) \propto \sqrt{1-\alpha_0^2}\sin\theta_{ee}\cos\theta_{ee}\sin\Delta_0,
\end{equation}
where $P_y$ shows the transverse polarization component that is perpendicular to the production plane. Here, the parameters $\alpha_0$ and $\Delta_0$, which relate to $\lambdacp\lambdacm$ production, are fixed according to the measurement reported in Ref.~\cite{BESIII:2025zbz}. We use the helicity amplitudes $H^{e^+e^-\to\lambdacp\lambdacm}_{\frac{1}{2},-\frac{1}{2}}$ and $H^{e^+e^-\to\lambdacp\lambdacm}_{\frac{1}{2},\frac{1}{2}}$ to express $\alpha_0$ and $\Delta_0$:
\begin{equation}
	\begin{aligned}
	\alpha_0 &= \frac{\left|H^{e^+e^-\to\lambdacp\lambdacm}_{\frac{1}{2},-\frac{1}{2}}\right|^2-2\left|H^{e^+e^-\to\lambdacp\lambdacm}_{\frac{1}{2},\frac{1}{2}}\right|^2}{\left|H^{e^+e^-\to\lambdacp\lambdacm}_{\frac{1}{2},-\frac{1}{2}}\right|^2+2\left|H^{e^+e^-\to\lambdacp\lambdacm}_{\frac{1}{2},\frac{1}{2}}\right|^2}, \\
	\Delta_0 &= \arg(H^{e^+e^-\to\lambdacp\lambdacm}_{\frac{1}{2},-\frac{1}{2}}) - \arg(H^{e^+e^-\to\lambdacp\lambdacm}_{\frac{1}{2},\frac{1}{2}}).
	\label{equ:alph0_delta0}
	\end{aligned}
\end{equation}

For simplicity, an overall amplitude for each component in Eq.~\eqref{eq:helicty13}, such as those listed in Table~\ref{tab:nominal1}, can be derived to represent the overall magnitude and phase of the specific component. In each cascade process, one of the partial-wave amplitudes $g_{ls}$ is taken as a reference with a fixed magnitude 1 and phase 0 rad. Table~\ref{tab:nominal2} lists the fixed amplitudes. Hence, the relative magnitudes and phases of the other partial-wave amplitudes in each cascade process are left free in the fit.

\subsection{Likelihood function construction and fit fraction}
\label{sec:likelihood}
\hspace{1.5em}
The signal PDF for a given event is constructed using the full amplitude as
\begin{eqnarray}
P = \frac{|\mathcal{A}|^2}{\int|\mathcal{A}|^2 \rm{d} \Phi}.
\label{eq:helicty14}
\end{eqnarray}
The integration of the denominator is calculated using an MC method based on sufficiently large PHSP MC samples of selected events, such that
\begin{eqnarray}
    \int |\mathcal{A}|^2 \mathrm{d}\Phi \propto \frac{1}{N_{\mathrm{PHSP}}} \sum_{i \in \mathrm{PHSP}} |\mathcal{A}(x_i)|^2.
\label{eq:helictyEx}
\end{eqnarray}

The negative log likelihood (NLL) is constructed by summing all signal candidates and subtracting the $M_{\rm miss}^2$ sideband backgrounds
\begin{eqnarray}
    -\ln L =  - \alpha \left[\sum_{i=1}^{N_{\rm data}} \ln P(x_i) - w'_{\text{bkg}} \sum_{j=1}^{N_{\rm sideband}} \ln P(x_j)\right],
\label{eq:helicty15}
\end{eqnarray}
where $w'_{\text{bkg}}=w_{\text{bkg}}\cdot\frac{N_{\text{data}}}{N_{\text{sideband}}}$, $w_{\text{bkg}}$ is the background fraction listed in Table~\ref{tab:fitmiss2}, $N_{\text{data}}$ and $N_{\text{sideband}}$ are the number of events in the $M_{\rm miss}^2$ signal region and sideband region, respectively. To achieve an unbiased uncertainty estimation, the normalization factor derived from Ref.~\cite{sWeight} is taken into account, which is expressed as
\begin{eqnarray}
    \alpha = \frac{N_{\text{data}} - N_{\text{sideband}}w'_{\text{bkg}}}{N_{\text{data}} + N_{\text{sideband}}{w'_{\text{bkg}}}^{2} }.
\label{eq:helicty16}
\end{eqnarray}
An individual NLL is first constructed separately for a given energy point and the joint NLL is obtained by summing over the NLL values of the different energy points.  After minimizing the joint NLL, the parameter covariance matrix is calculated from the inverse Hessian matrix
\begin{eqnarray}
    V_{ij}^{-1} = - \frac{\partial^2 \ln L}{\partial X_i \partial X_j},
\label{eq:helicty18}
\end{eqnarray}
where $X_i$ is the $i$-th free parameter in the fit. 

The fit fraction (FF) for each resonant component can be calculated as
\begin{eqnarray}
    \mathrm{FF}_i = \frac{\int |\mathcal{A}_i|^2 \mathrm{d} \Phi'}
    {\int |\sum_k \mathcal{A}_k|^2 \mathrm{d} \Phi'},
\label{eq:FF}
\end{eqnarray}
where $\mathcal{A}_i$ is the amplitude of the $i$-th component and the integration is calculated by the sum of truth level PHSP MC samples before requiring detector acceptance. Hence, the FFs of the interference part can be calculated as
\begin{eqnarray}
    \mathrm{FF}_{i,j} = \frac{\int |\mathcal{A}_i+\mathcal{A}_j|^2 \mathrm{d} \Phi'}
    {\int |\sum_k \mathcal{A}_k|^2 \mathrm{d} \Phi'}-\mathrm{FF}_i-\mathrm{FF}_j.
\label{eq:interference}
\end{eqnarray}
The FFs statistical uncertainties are obtained accounting for correlations.

\subsection{Fit results}
\label{sec:nominal}
\hspace{1.5em}

The established excited $\Delta^{*+}$, $N^{*+}$ and $\Lambda^{*0}$ \cite{ParticleDataGroup:2024cfk} are included in the amplitude model, then statistical significances are evaluated including the following amplitudes: $\Delta(1232)^{+}$, $\Delta(1600)^{+}$, $\Delta(1620)^{+}$, $\Delta(1700)^{+}$, $N(1440)^{+}$, $N(1520)^{+}$, $N(1535)^{+}$, $N(1650)^{+}$, $N(1675)^{+}$, $N(1680)^{+}$, $N(1710)^{+}$, $\Lambda(1405)^{0}$, $\Lambda(1520)^{0}$, $\Lambda(1600)^{0}$, $\Lambda(1670)^{0}$, $\Lambda(1690)^{0}$, $\Lambda(1830)^{0}$, $\Lambda(1890)^{0}$ and $\Lambda(2000)^{0}$, $\mathcal{S}$-wave $(N\!R_{1/2^{-}})$ and $\mathcal{P}$-wave $(N\!R_{1/2^{+}}, N\!R_{3/2^{+}})$. The statistical significance of each component is calculated from the difference of the NLL values between fits with and without the component, accounting for the corresponding change in the number of degrees of freedom. The baseline model includes amplitudes with significance greater than $3\sigma$. These components are $\Lambda(1830)^{0}$, $\Lambda(2000)^{0}$, $N(1440)^{+}$ and $N(1650)^{+}$. The mass and width parameters of the $\Lambda(1830)^{0}$, $N(1440)^{+}$ and $N(1650)^{+}$ states are fixed to the world-average values~\cite{ParticleDataGroup:2024cfk} and those of the $\Lambda(2000)^{0}$ state are taken from the most recent measurements~\cite{Sarantsev:2019xxm}. Neither a narrow $N(1535)^+$ peak nor a dip induced by $\Lambda(1670)^0$ is observed in the invariant-mass spectra of $n\pip$ and $n\Ks$. Moreover, although the $\Lambda(1670)^0$ state has a significance of $2.6\sigma$, it appears as a narrow peak near $1.7\,\gevcc$ in the $n\Ks$ invariant-mass spectrum rather than as a dip structure.

The NLL of the baseline fit contains fourteen free parameters in total, consisting of magnitudes and phases of the total amplitudes (six parameters) and partial-wave amplitudes (eight parameters). Tables~\ref{tab:nominal1} and \ref{tab:nominal2} list the fit results.  Table~\ref{tab:interference} lists the interference fractions. Figures~\ref{fig:nominal1} and \ref{fig:nominal2} show one-dimensional projections of the invariant masses and the helicity angles, respectively. We obtain the product BFs of the resonances by multiplying the FFs listed in Table~\ref{tab:nominal1} with the total three-body BF $\BR(\sigmode)=(1.86\pm 0.08\pm0.04)\%$~\cite{BESIII:2023pia}. Table~\ref{tab:BF_alpha} lists the absolute BFs.

\begin{table*}[!tp]
\caption{Total amplitudes for different components in the baseline fit, along with their FFs and significances. The total FF of intermediate resonances is 94.8\%. Only statistical uncertainties are listed.}
\setlength{\abovecaptionskip}{1.2cm}
\setlength{\belowcaptionskip}{0.2cm}
\label{tab:nominal1}
\begin{center}
\vspace{-0.0cm}
\begin{tabular}{c  c c c c}
	\hline \hline
	Process & Magnitude & Phase $\phi$ (rad) & FF ($\%$) & Significance \\
	\hline
    $N(1650)^+\Ks$     & $ 1 $ (fixed)   & $ 0 $   (fixed)  & $26.0\pm5.4\phantom{0}$  & $4.8\,\sigma$\\
    $N(1440)^+\Ks$     & $1.36\pm0.32$  & $-0.78\pm0.63$  & $25.9\pm10.4$  & $5.2\,\sigma$\\
    $\Lambda(1830)^0\pi^+$    & $0.65\pm0.29$  & $-3.19\pm0.47$  & $28.4\pm9.7\phantom{0}$ & $5.5\,\sigma$\\
    $\Lambda(2000)^0\pi^+$    & $0.96\pm0.25$ & $-1.68\pm0.59$ & $ 14.5\pm4.1\phantom{0}$  & $4.5\,\sigma$\\
	\hline\hline
\end{tabular}
\end{center}
\end{table*}
\vspace{-0.0cm}

\renewcommand{\arraystretch}{1.5}

\begin{table*}[!tp]
\caption{Partial wave amplitudes $g_{ls}$ for different resonances in the baseline fit. Only statistical uncertainties are listed.}  \setlength{\abovecaptionskip}{1.2cm}
\setlength{\belowcaptionskip}{0.2cm}
\label{tab:nominal2}
\begin{center}
\vspace{-0.0cm}
\begin{tabular}{c c c | c c c}
	\hline \hline
 	Amplitude & Magnitude & Phase $\phi$ (rad)
	& Amplitude & Magnitude & Phase $\phi$ (rad)\\
    \hline
	\multicolumn{3}{c|}{$\frac{1}{2}^{+}(\lcp)\to\frac{1}{2}^{+}(N(1440)^+)+0^-(\Ks)$} 
	& \multicolumn{3}{c}{$\frac{1}{2}^{+}(\lcp)\to\frac{1}{2}^{-}(N(1650)^+)+0^-(\Ks)$}\\
	\hline
	$g^{N(1440)^+}_{0\frac{1}{2}}$   & $1$ (fixed) & $0$ (fixed)
	& $g^{N(1650)^+}_{0\frac{1}{2}}$ & $1$ (fixed) & $0$ (fixed) \\
	$g^{N(1440)^+}_{1\frac{1}{2}}$   & $ -0.10 \pm 0.26 $ & $ -0.42 \pm 2.61 $
	& $g^{N(1650)^+}_{1\frac{1}{2}}$ & $ 0.41 \pm 0.27 $ & $ 2.05 \pm 0.70 $\\
	\hline\hline
	
	\multicolumn{3}{c|}{$\frac{1}{2}^{+}(\lcp)\to\frac{5}{2}^{-}(\Lambda(1830)^0)+0^-(\pip)$} 
	& \multicolumn{3}{c}{$\frac{1}{2}^{+}(\lcp)\to\frac{1}{2}^{-}(\Lambda(2000)^0)+0^-(\pip)$}\\
	\hline
	$g^{\Lambda(1830)^0}_{2\frac{5}{2}}$   & $1$ (fixed)  & $0$ (fixed)
	& $g^{\Lambda(2000)^0}_{0\frac{1}{2}}$ & $1$ (fixed)  & $0$ (fixed) \\
	$g^{\Lambda(1830)^0}_{3\frac{5}{2}}$   & $ 1.11 \pm 0.70 $ & $ 0.35 \pm 0.76 $
	& $g^{\Lambda(2000)^0}_{1\frac{1}{2}}$ & $ 1.40 \pm 0.77 $ & $ -3.41 \pm 0.80 $\\
	\hline\hline
\end{tabular}
\end{center}
\end{table*}
\vspace{-0.0cm}
\renewcommand{\arraystretch}{1}

\begin{table*}[!htbp]
\caption{FFs of the interference terms between different intermediate resonant processes in the baseline fit.  Only statistical uncertainties are listed.}
\setlength{\abovecaptionskip}{1.2cm}
\setlength{\belowcaptionskip}{0.2cm}
\label{tab:interference}
\begin{center}
\vspace{-0.0cm}
\begin{tabular}{c | c c c}
	\hline \hline
	$\mathrm{FF}_{i,j}(\%)$ & $\Lambda(1830)^0\pip$ & $\Lambda(2000)^0\pip$ & $N(1440)^+\Ks$ \\
	\hline
    $\Lambda(2000)^0\pip$ & $ 0.0 \pm 0.0 $ \\
    $N(1440)^+\Ks$ & $ 0.8 \pm 0.6 $ & $ \phantom{-}6.3 \pm 3.6 $ \\
    $N(1650)^+\Ks$ & $ 0.3 \pm 0.3 $ & $ -2.1 \pm 4.4 $ & $ 0.0 \pm 0.0 $  \\
	\hline\hline
\end{tabular}
\end{center}
\end{table*}
\vspace{-0.0cm}


\vspace{-0.0cm}
\begin{figure}[t]
\setlength{\abovecaptionskip}{-1pt}
\setlength{\belowcaptionskip}{10pt}
\centering
\includegraphics[trim = 9mm 0mm 0mm 0mm, width=0.45\textwidth]{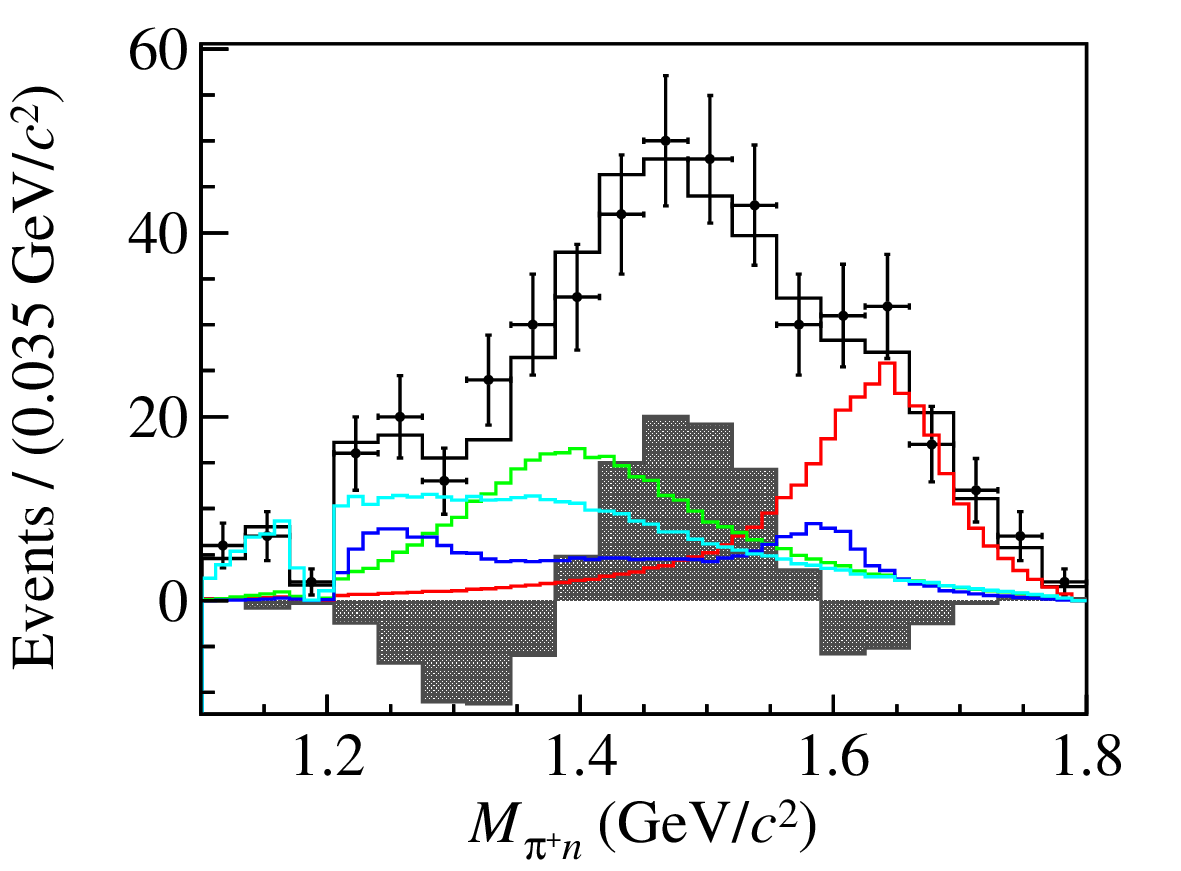}
\includegraphics[trim = 9mm 0mm 0mm 0mm, width=0.45\textwidth]{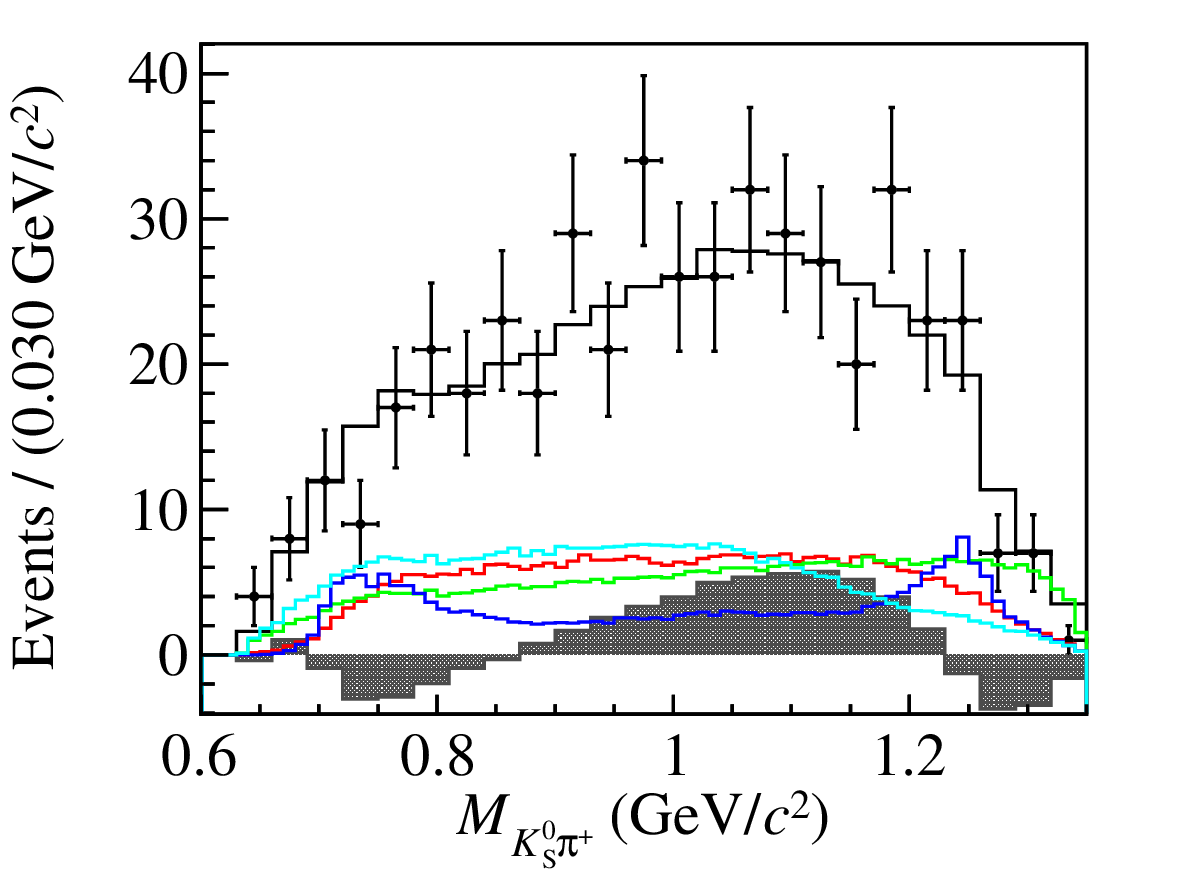}\\
\includegraphics[trim = 9mm 0mm 0mm 0mm, width=0.45\textwidth]{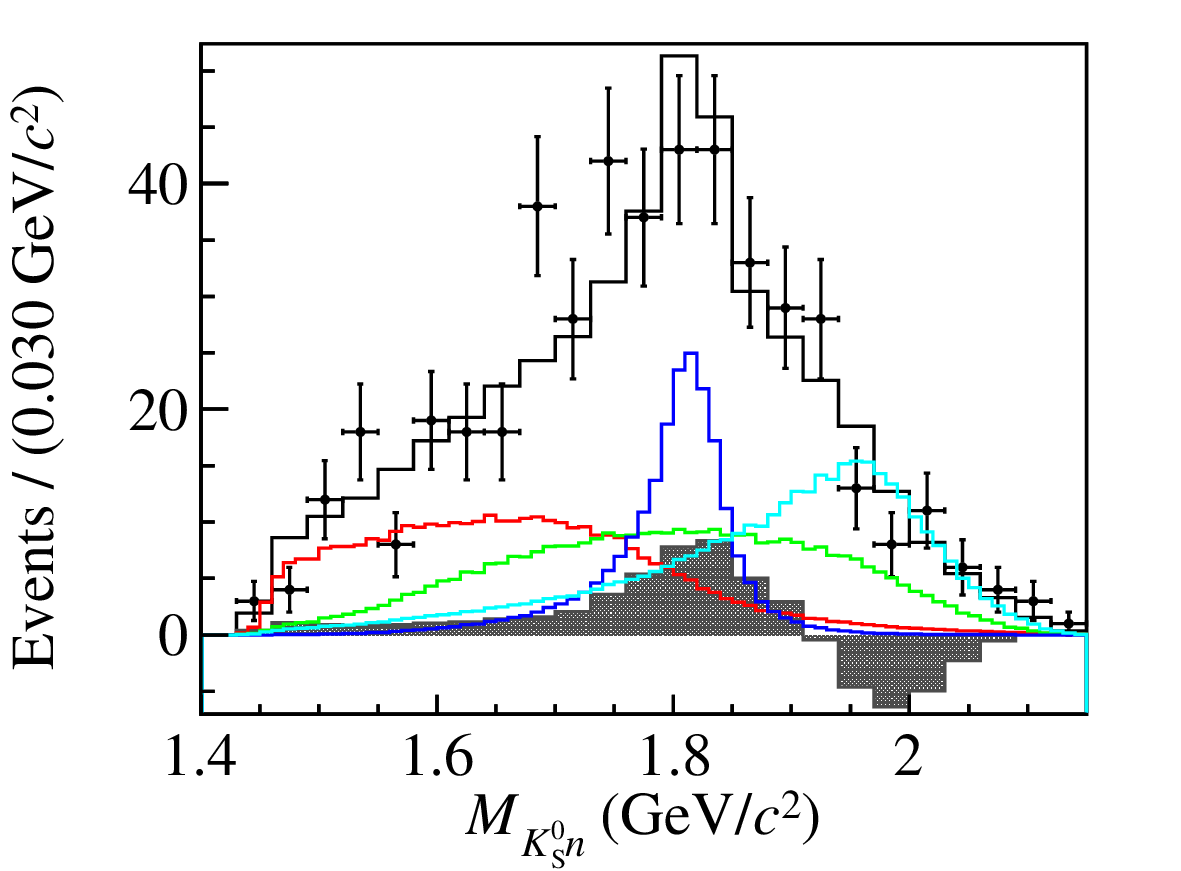}
\includegraphics[trim = 0mm 0mm 0mm 0mm, width=0.24\textwidth]{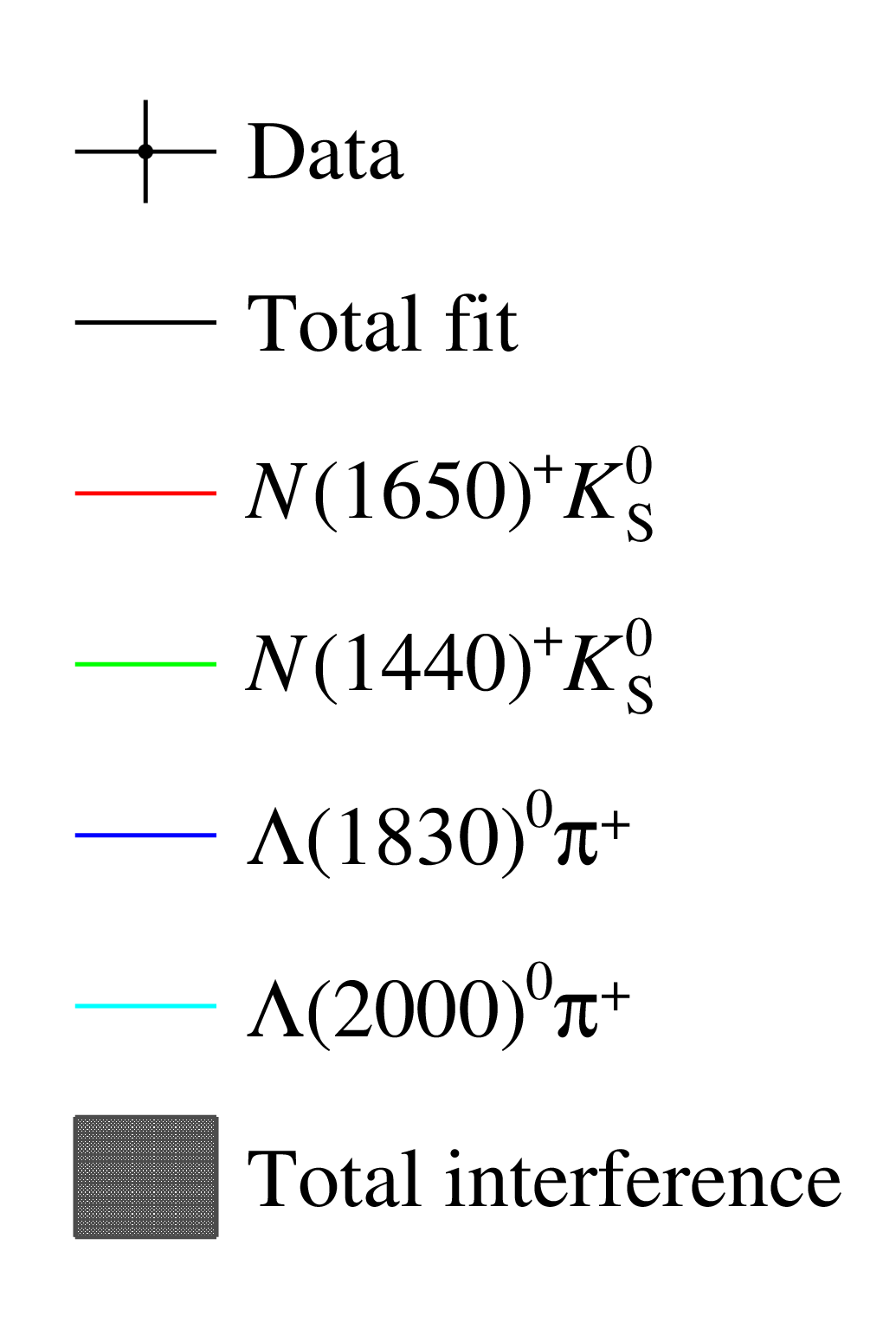}
\caption{Projections of the fit results in the invariant mass spectra
  $M_{\pip n}$, $M_{\Ks\pip}$ and $M_{\Ks n}$.}
\label{fig:nominal1}
\end{figure}

\vspace{-0.0cm}
\begin{figure}[thbp]
\setlength{\abovecaptionskip}{-1pt}
\setlength{\belowcaptionskip}{10pt}
\centering
\includegraphics[trim = 9mm 0mm 0mm 0mm, width=0.32\textwidth]{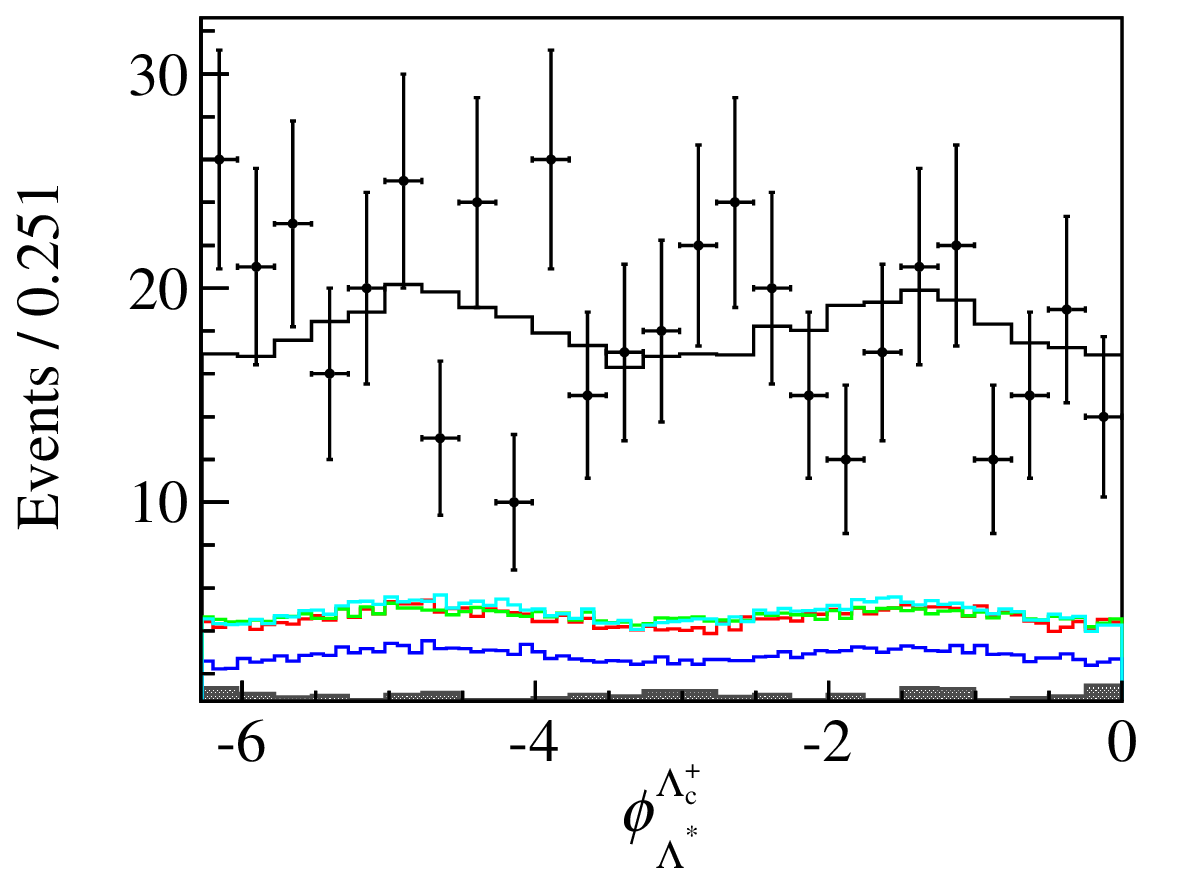}
\includegraphics[trim = 9mm 0mm 0mm 0mm, width=0.32\textwidth]{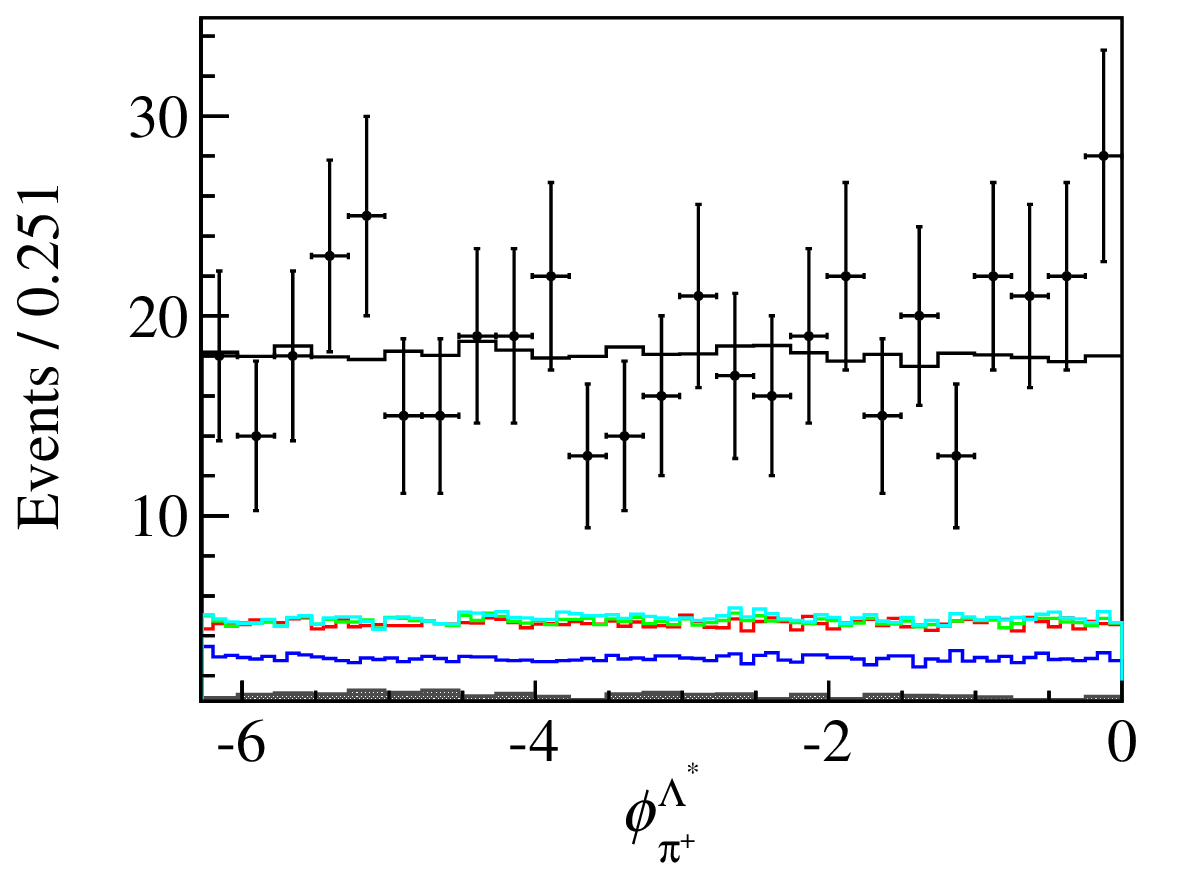}
\includegraphics[trim = 9mm 0mm 0mm 0mm, width=0.32\textwidth]{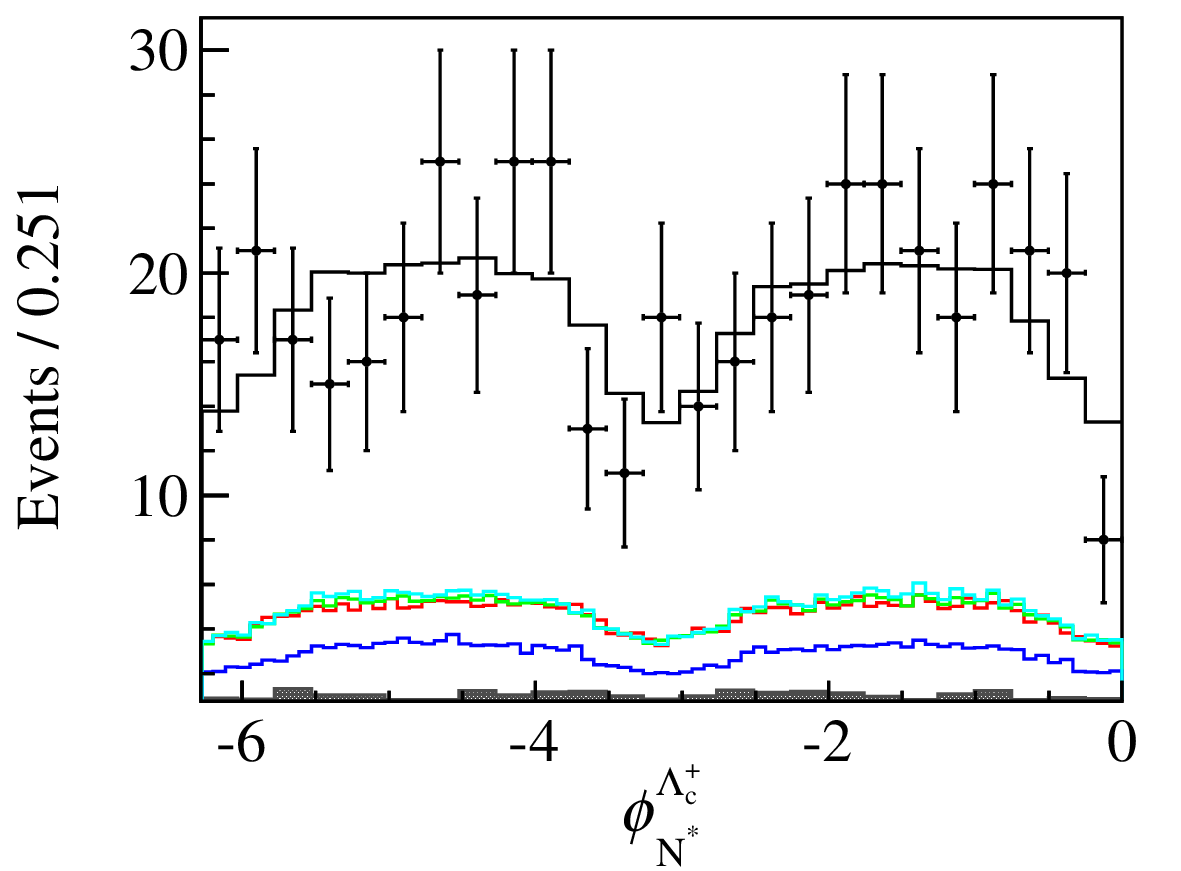}\\
\includegraphics[trim = 9mm 0mm 0mm 0mm, width=0.32\textwidth]{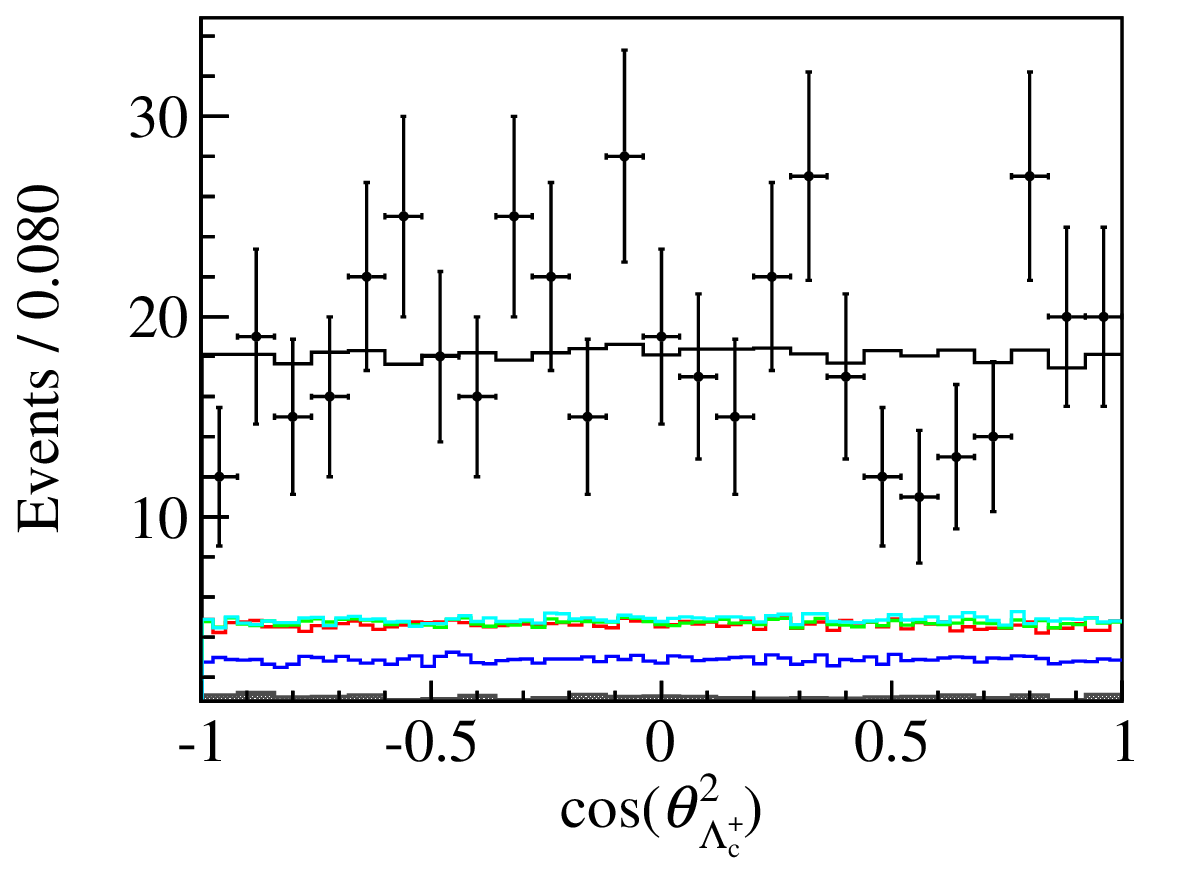}
\includegraphics[trim = 9mm 0mm 0mm 0mm, width=0.32\textwidth]{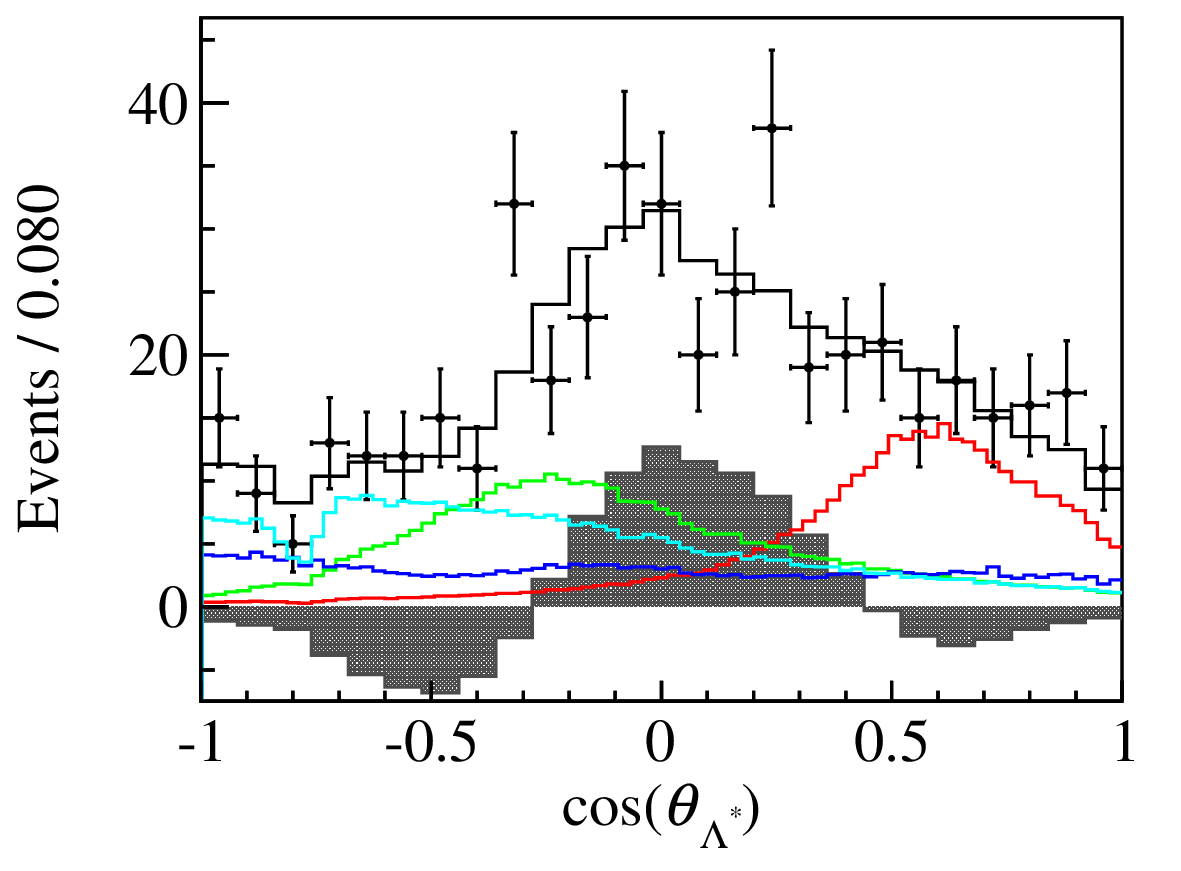}
\includegraphics[trim = 9mm 0mm 0mm 0mm, width=0.32\textwidth]{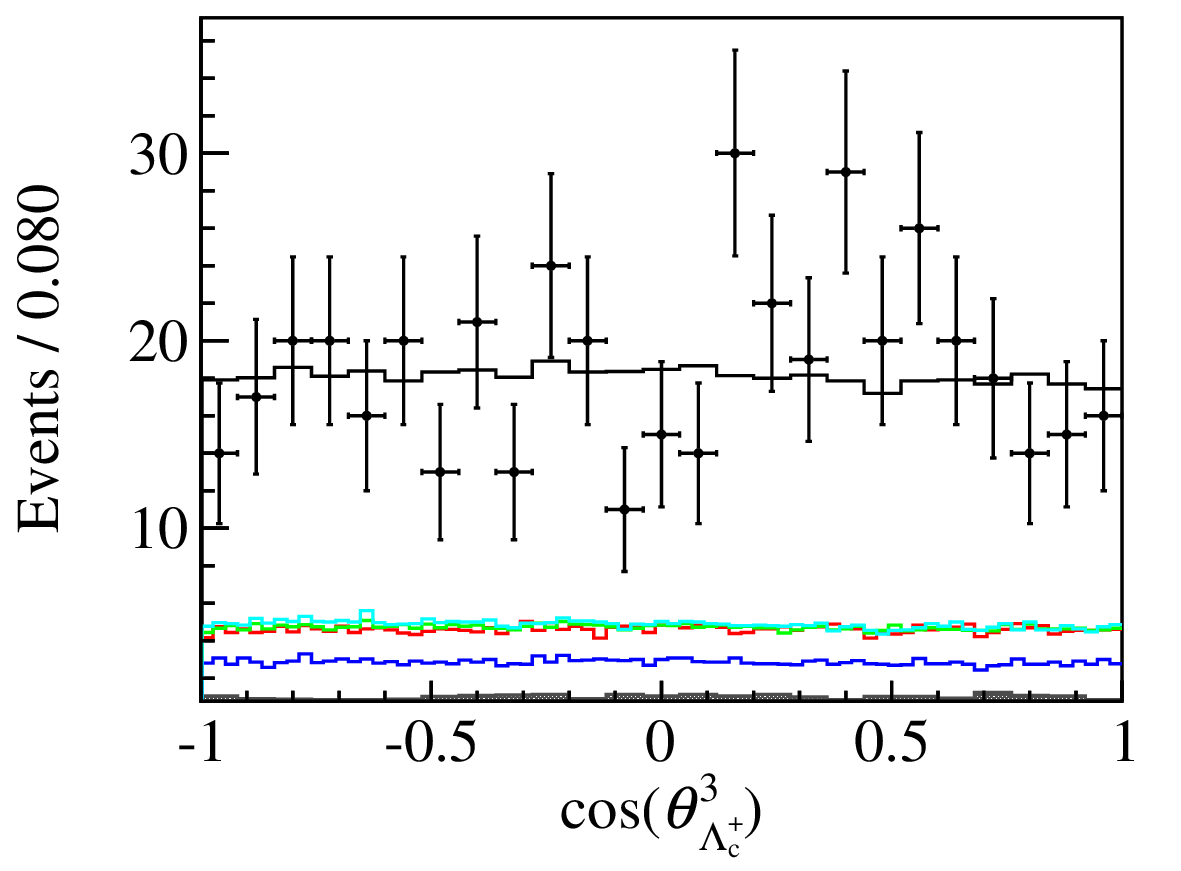}\\
\includegraphics[trim = 9mm 0mm 0mm 0mm, width=0.32\textwidth]{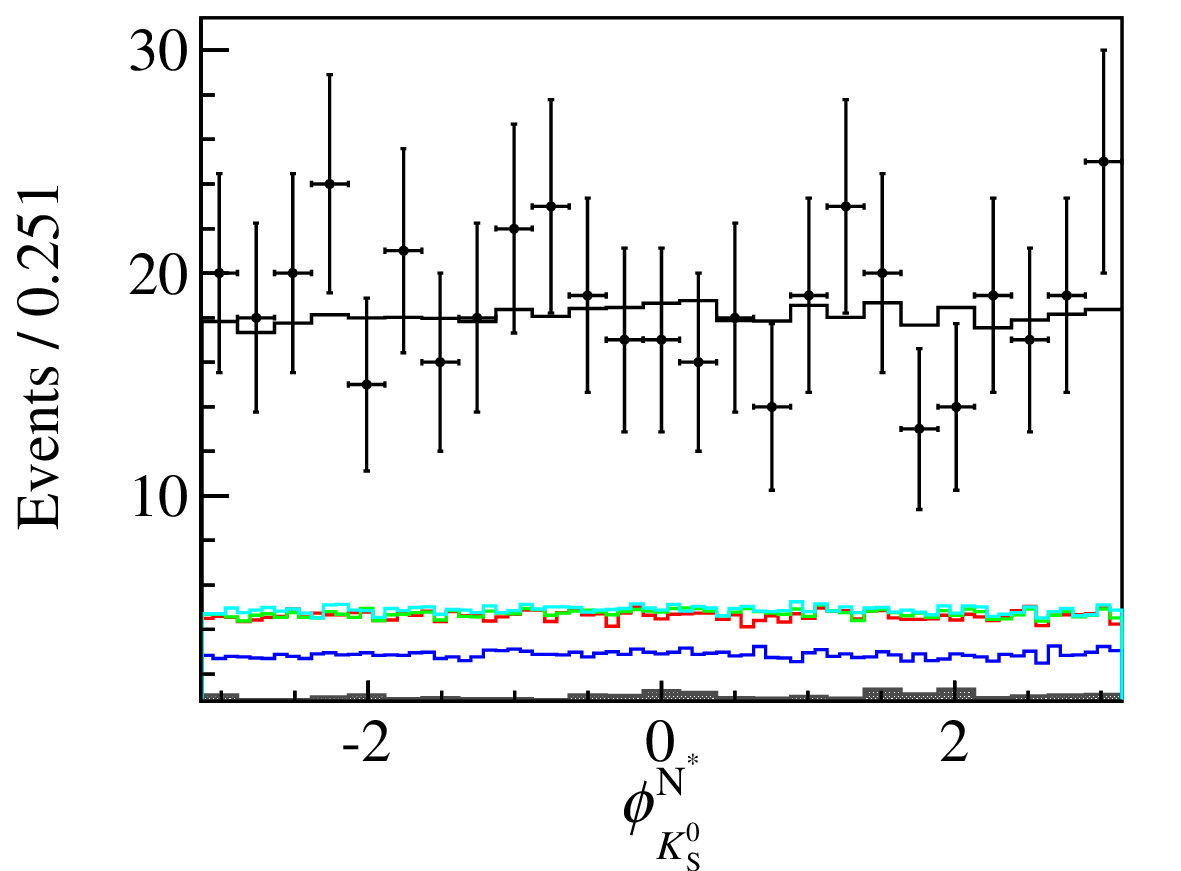}
\includegraphics[trim = 9mm 0mm 0mm 0mm, width=0.32\textwidth]{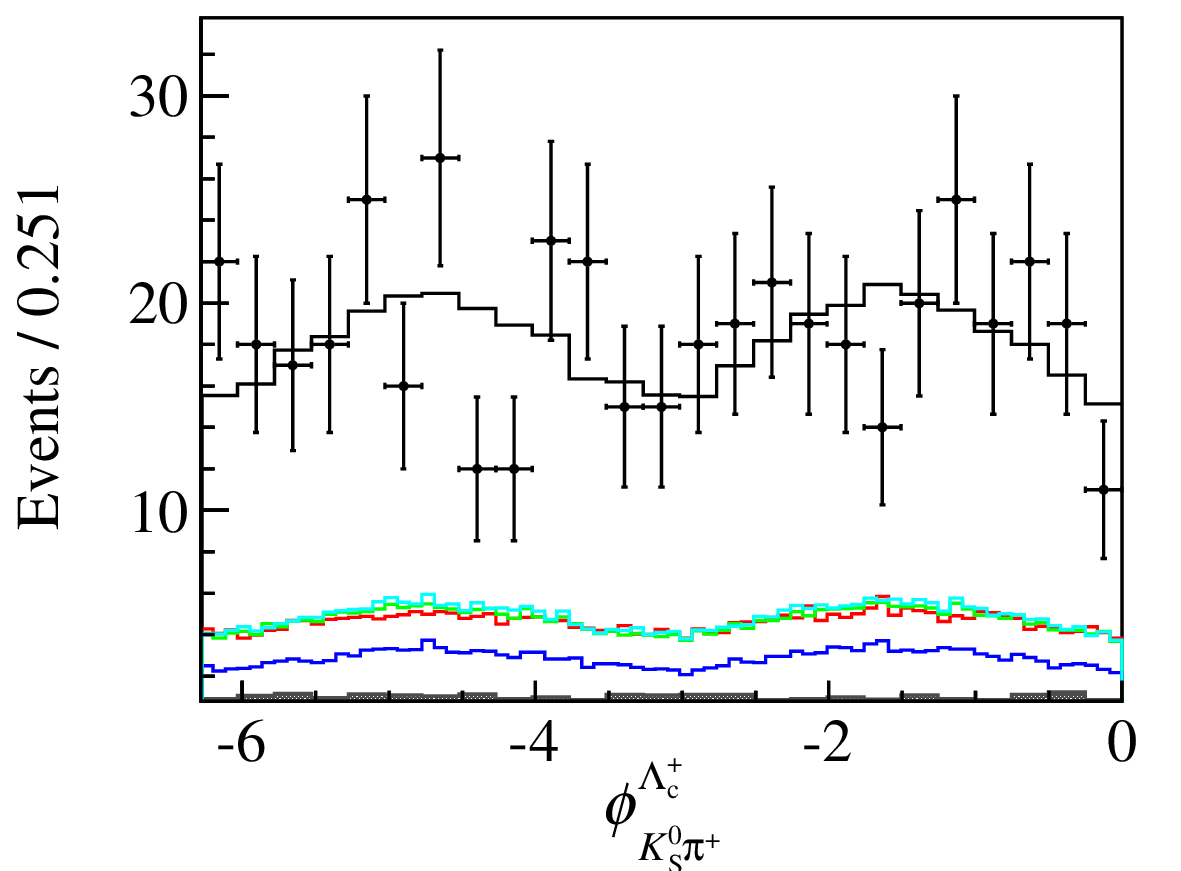}
\includegraphics[trim = 9mm 0mm 0mm 0mm, width=0.32\textwidth]{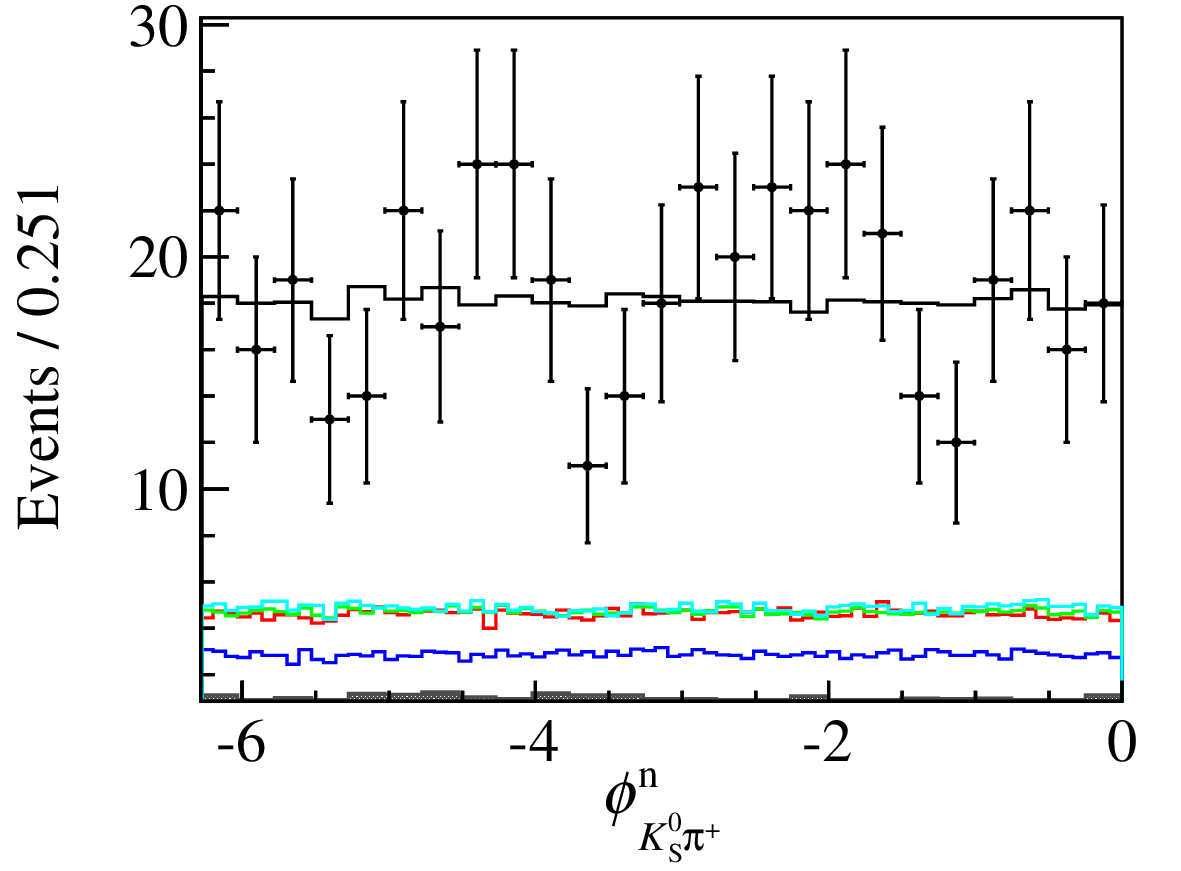}\\
\includegraphics[trim = 9mm 0mm 0mm 0mm, width=0.32\textwidth]{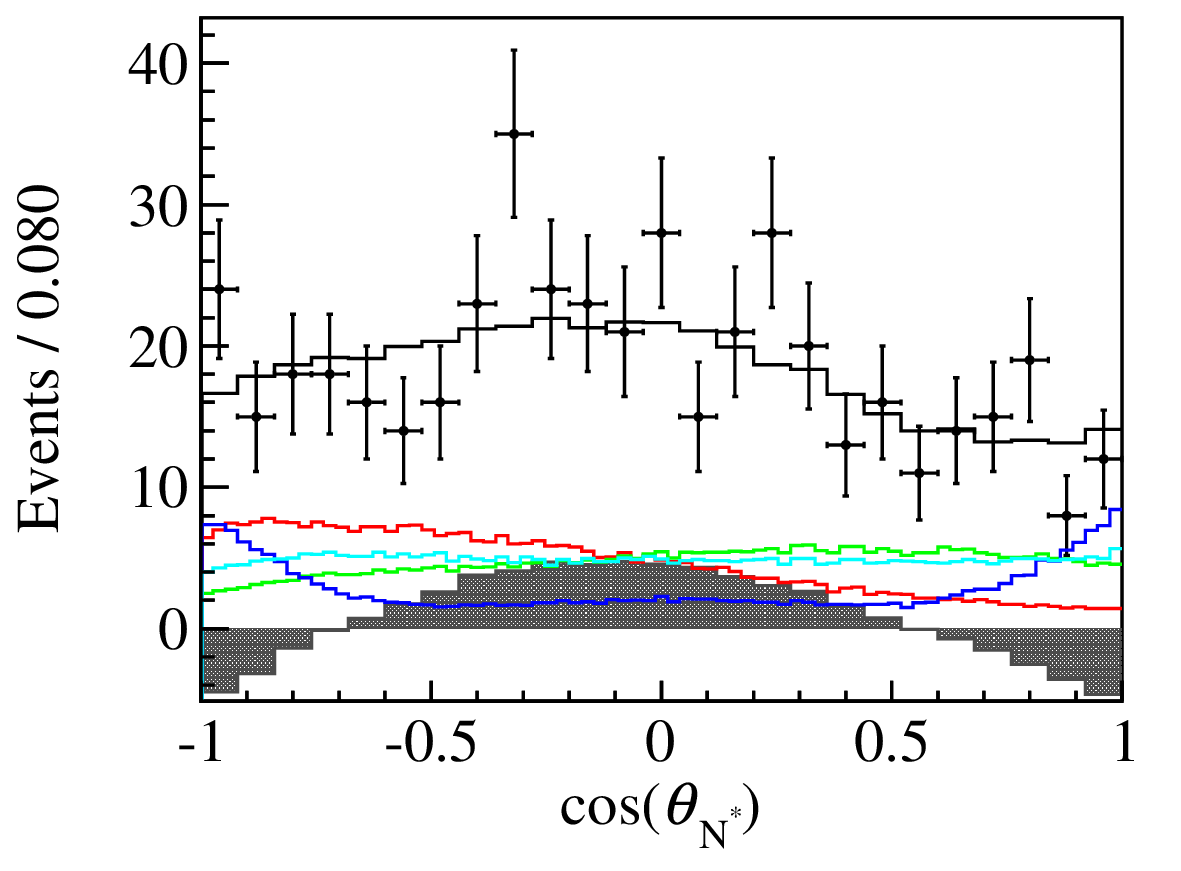}
\includegraphics[trim = 9mm 0mm 0mm 0mm, width=0.32\textwidth]{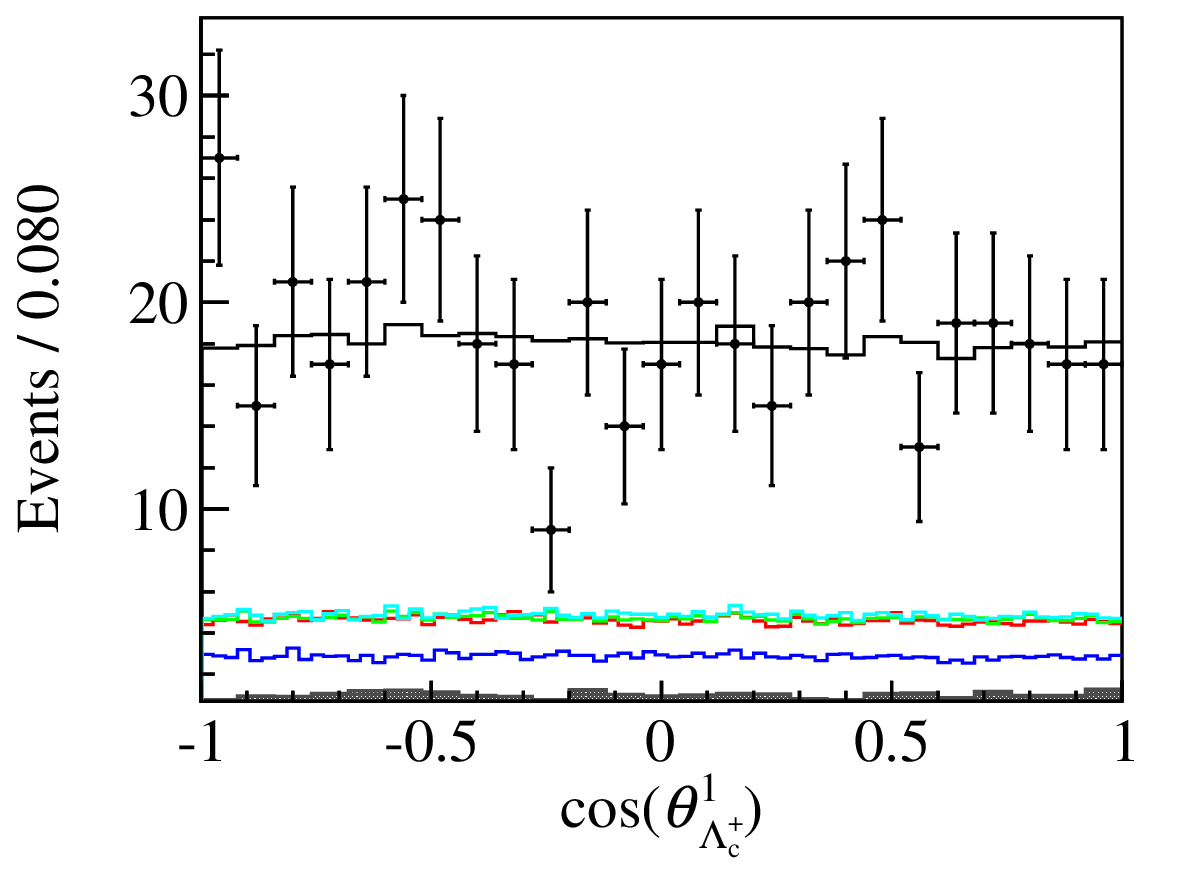}
\includegraphics[trim = 9mm 0mm 0mm 0mm, width=0.32\textwidth]{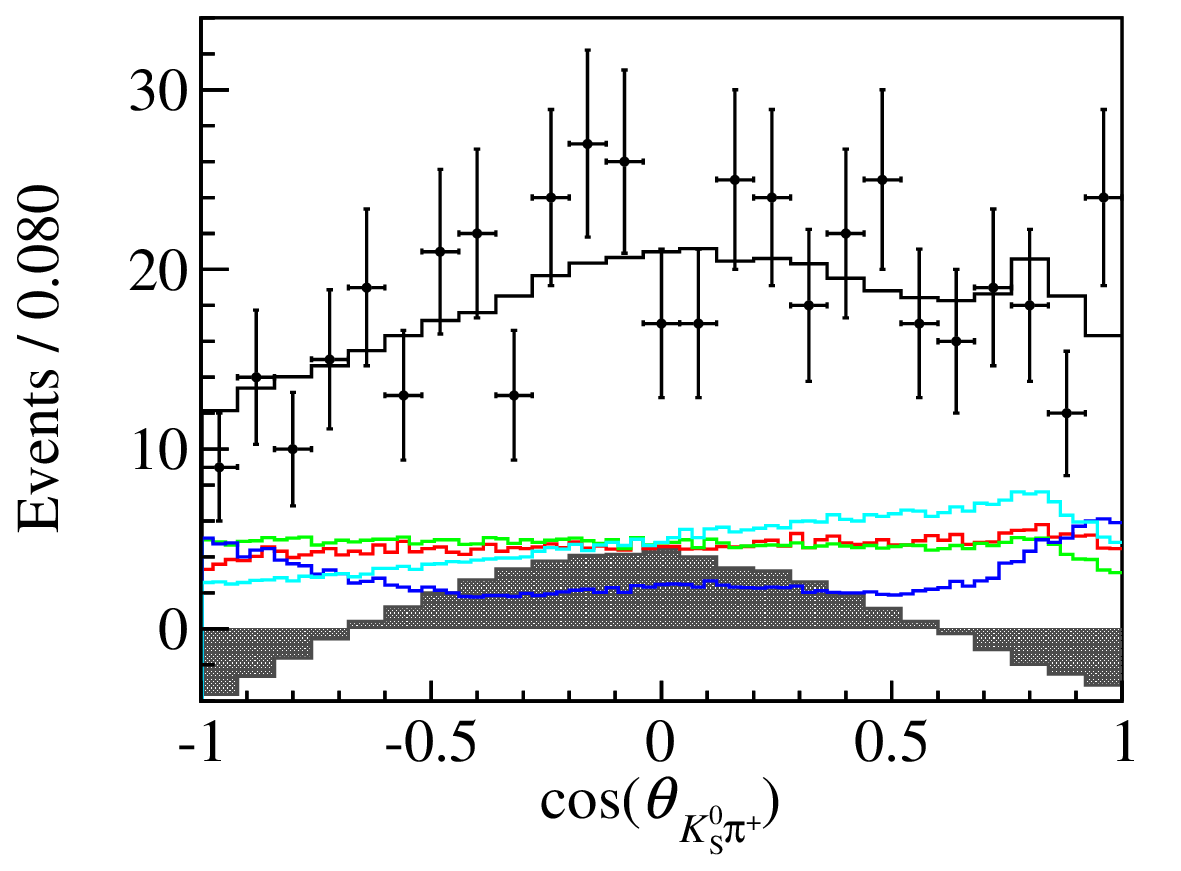}\\
\caption{Projections of the fit results in the helicity angle distributions. The legend of components is shown in Fig.~\ref{fig:nominal1}}
\label{fig:nominal2}
\end{figure}

\clearpage
\subsection{Decay-asymmetry parameters}
\label{sec:asymmetry}
\hspace{1.5em} The decay asymmetry parameters $\alpha$ are related to the interference effects among the different partial waves. When the intermediate state is at its pole mass, the helicity amplitude of a two-body decay can be written under the LS coupling expansion as
\begin{eqnarray}
    H_{\lambda_1, \lambda_2} = \sum_{l s} g_{ls} \sqrt{\frac{2l+1}{2J_0 + 1}} 
    \langle l 0s  \delta| J_0\delta \rangle
    \langle J_1 \lambda_1 J_2\ -\lambda_2| s\delta \rangle.
\label{eq:ls2}
\end{eqnarray}
The helicity amplitudes of four different intermediate processes are
\begin{equation}
\begin{aligned}
H^{\lambdacp\to\Lambda(1830)^{0}\pip}_{\mp\frac{1}{2},0} &= \frac{\left(g^{\lambdacp\to\Lambda(1830)^{0}\pip}_{2,\frac{5}{2}}\pm g^{\lambdacp\to\Lambda(1830)^{0}\pip}_{3,\frac{5}{2}}\right)}{\sqrt{2}}, \\
H^{\lambdacp\to\Lambda(2000)^{0}\pip}_{\mp\frac{1}{2},0} &= \frac{\left(g^{\lambdacp\to\Lambda(2000)^{0}\pip}_{0,\frac{1}{2}}\pm g^{\lambdacp\to\Lambda(2000)^{0}\pip}_{1,\frac{1}{2}}\right)}{\sqrt{2}}, \\
H^{\lambdacp\to N(1440)^{+}\Ks}_{\mp\frac{1}{2},0} &= \frac{\left(g^{\lambdacp\to N(1440)^{+}\Ks}_{0,\frac{1}{2}}\pm g^{\lambdacp\to N(1440)^{+}\Ks}_{1,\frac{1}{2}}\right)}{\sqrt{2}}, \\
H^{\lambdacp\to N(1650)^{+}\Ks}_{\mp\frac{1}{2},0} &= \frac{\left(g^{\lambdacp\to N(1650)^{+}\Ks}_{0,\frac{1}{2}}\pm g^{\lambdacp\to N(1650)^{+}\Ks}_{1,\frac{1}{2}}\right)}{\sqrt{2}}. \\
\label{equ:LS_nominal_mass}
\end{aligned}
\end{equation}
Thus, the decay asymmetries expressed in terms of the partial-wave amplitudes are
\begin{equation*}
\begin{aligned}
\alpha_{\Lambda(1830)^{0}\pi^{+}} &=\frac{\left|H_{\frac{1}{2}, 0}^{\lambdacp\to\Lambda(1830)^{0}\pi^{+}}\right|^2-\left|H_{-\frac{1}{2}, 0}^{\lambdacp\to\Lambda(1830)^{0}\pi^{+}}\right|^2}{\left|H_{\frac{1}{2}, 0}^{\lambdacp\to\Lambda(1830)^{0}\pi^{+}}\right|^2+\left|H_{-\frac{1}{2}, 0}^{\lambdacp\to\Lambda(1830)^{0}\pi^{+}}\right|^2}=\frac{2 \Re\left(g_{2, \frac{5}{2}}^{\lambdacp\to\Lambda(1830)^{0}\pi^{+}} \cdot g_{3, \frac{5}{2}}^{\lambdacp\to\Lambda(1830)^{0}\pi^{+}*}\right)}{\left|g_{2, \frac{5}{2}}^{\lambdacp\to\Lambda(1830)^{0}\pi^{+}}\right|^2+\left|g_{3, \frac{5}{2}}^{\lambdacp\to\Lambda(1830)^{0}\pi^{+}}\right|^2}, \\
\alpha_{\Lambda(2000)^{0}\pi^{+}} &=\frac{\left|H_{\frac{1}{2}, 0}^{\lambdacp\to\Lambda(2000)^{0}\pi^{+}}\right|^2-\left|H_{-\frac{1}{2}, 0}^{\lambdacp\to\Lambda(2000)^{0}\pi^{+}}\right|^2}{\left|H_{\frac{1}{2}, 0}^{\lambdacp\to\Lambda(2000)^{0}\pi^{+}}\right|^2+\left|H_{-\frac{1}{2}, 0}^{\lambdacp\to\Lambda(2000)^{0}\pi^{+}}\right|^2}=\frac{2 \Re\left(g_{0, \frac{1}{2}}^{\lambdacp\to\Lambda(2000)^{0}\pi^{+}} \cdot g_{1, \frac{1}{2}}^{\lambdacp\to\Lambda(2000)^{0}\pi^{+}*}\right)}{\left|g_{0, \frac{1}{2}}^{\lambdacp\to\Lambda(2000)^{0}\pi^{+}}\right|^2+\left|g_{1, \frac{1}{2}}^{\lambdacp\to\Lambda(2000)^{0}\pi^{+}}\right|^2}, \\
\alpha_{N(1440)^{+}\Ks} &=\frac{\left|H_{\frac{1}{2}, 0}^{\lambdacp\to N(1440)^{+}\Ks}\right|^2-\left|H_{-\frac{1}{2}, 0}^{\lambdacp\to N(1440)^{+}\Ks}\right|^2}{\left|H_{\frac{1}{2}, 0}^{\lambdacp\to N(1440)^{+}\Ks}\right|^2+\left|H_{-\frac{1}{2}, 0}^{\lambdacp\to N(1440)^{+}\Ks}\right|^2}=\frac{2 \Re\left(g_{0, \frac{1}{2}}^{\lambdacp\to N(1440)^{+}\Ks} \cdot g_{1, \frac{1}{2}}^{\lambdacp\to N(1440)^{+}\Ks*}\right)}{\left|g_{0, \frac{1}{2}}^{\lambdacp\to N(1440)^{+}\Ks}\right|^2+\left|g_{1, \frac{1}{2}}^{\lambdacp\to N(1440)^{+}\Ks}\right|^2}, \\
\alpha_{N(1650)^{+}\Ks} &=\frac{\left|H_{\frac{1}{2}, 0}^{\lambdacp\to N(1650)^{+}\Ks}\right|^2-\left|H_{-\frac{1}{2}, 0}^{\lambdacp\to N(1650)^{+}\Ks}\right|^2}{\left|H_{\frac{1}{2}, 0}^{\lambdacp\to N(1650)^{+}\Ks}\right|^2+\left|H_{-\frac{1}{2}, 0}^{\lambdacp\to N(1650)^{+}\Ks}\right|^2}=\frac{2 \Re\left(g_{0, \frac{1}{2}}^{\lambdacp\to N(1650)^{+}\Ks} \cdot g_{1, \frac{1}{2}}^{\lambdacp\to N(1650)^{+}\Ks*}\right)}{\left|g_{0, \frac{1}{2}}^{\lambdacp\to N(1650)^{+}\Ks}\right|^2+\left|g_{1, \frac{1}{2}}^{\lambdacp\to N(1650)^{+}\Ks}\right|^2}. \\
\end{aligned}
\end{equation*}

\begin{figure}[htbp]
    \centering
    \includegraphics[width=0.45\textwidth]{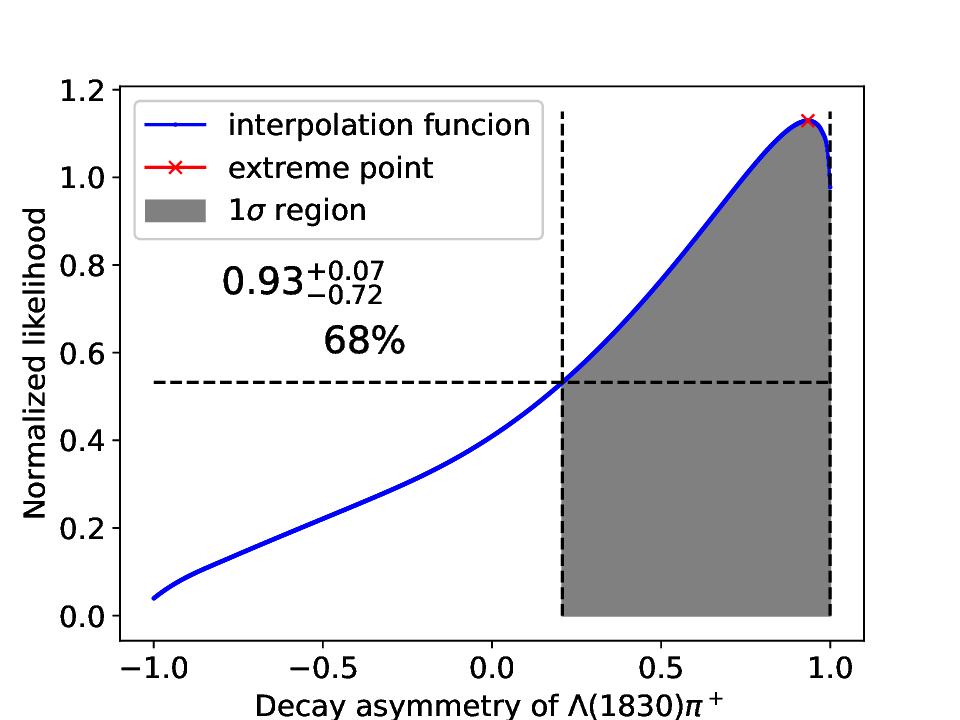}
    \includegraphics[width=0.45\textwidth]{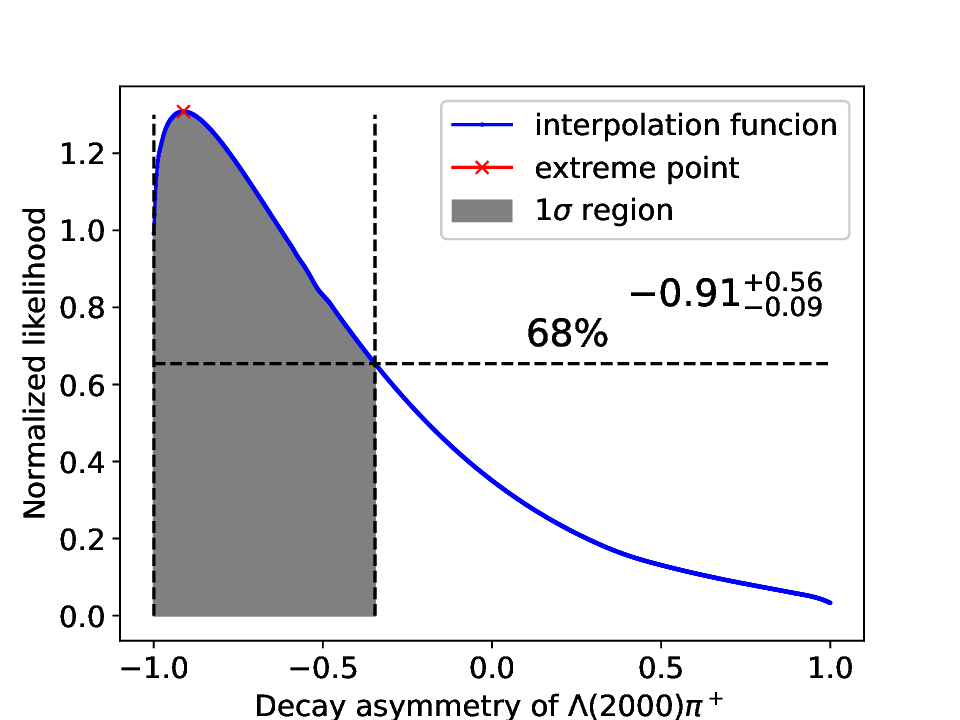}
\caption{Normalized likelihood distributions as a function of the decay asymmetries $\alpha_{\Lambda(1830)^0\pip}$ (left) and $\alpha_{\Lambda(2000)^0\pip}$ (right). The shaded regions denote the 68\% confidence intervals.}
\label{fig:alpha_NLL_scan}
\end{figure}

Using the baseline-fit results of the partial-wave amplitudes listed in Table~\ref{tab:nominal2}, we obtain the decay-asymmetry parameters listed in Table~\ref{tab:BF_alpha}. The statistical uncertainties in $\alpha_{N(1440)^+\Ks}$ and $\alpha_{N(1650)^+\Ks}$ are calculated using error propagation. However, for $\alpha_{\Lambda(1830)^0\pip}$ and $\alpha_{\Lambda(2000)^0\pip}$, the statistical uncertainties are estimated from NLL scans because the central values are close to the physical boundaries, leading to non-Gaussian NLL distributions. Figure~\ref{fig:alpha_NLL_scan} shows the normalized likelihood distributions, with the shaded regions indicating the $68\%$ confidence intervals. The asymmetric $1\sigma$ uncertainties are obtained from these intervals.

\begin{table*}[!htbp]
\caption{Decay asymmetry parameters after correction for fit bias as described in Sec.~\ref{sec:sys}. Uncertainties are statistical only.}
\setlength{\abovecaptionskip}{1.2cm}
\setlength{\belowcaptionskip}{0.2cm}
\label{tab:BF_alpha}
\begin{center}
\vspace{-0.0cm}
\begin{tabular}{c | c}
	\hline \hline
	 & Result \\
	\hline
    $\alpha_{\Lambda(1830)^0\pip}$ & $\phantom{-}0.93^{+0.07}_{-0.72}$ \\
    $\alpha_{\Lambda(2000)^0\pip}$ & $-0.91^{+0.56}_{-0.09}$ \\
    $\alpha_{N(1440)^+\Ks}$ & $-0.13\pm0.43$ \\
    $\alpha_{N(1650)^+\Ks}$ & $-0.42\pm0.45$ \\
	\hline\hline
\end{tabular}
\end{center}
\end{table*}
\vspace{-0.0cm}

\subsection{Upper limit of $\BR(\lcp \to \Delta(1232)^{+} \Bar{K}^0)$}
\label{sec:upper_limit}

\hspace{1.5em} The LHCb Collaboration published a partial-wave analysis of $\lcp \to p K^- \pip$ decay~\cite{LHCb:2022sck}. The results indicate a significant contribution from the intermediate resonant process $\Delta(1232)^{++} K^-$ that accounts for $(28.6\pm0.83)\%$ of the total rate. As the isospin-symmetric decay partner to $\lcp \to p K^- \pip$, the amplitude analysis performed in the present paper provides information about the analogous intermediate process $\Delta(1232)^{+} \Bar{K}^0$, allowing a comparison of the decay dynamics of the two channels.

\begin{figure*}[thbp]
\centering
\subfigure[]
{
\includegraphics[width=0.3\textwidth]{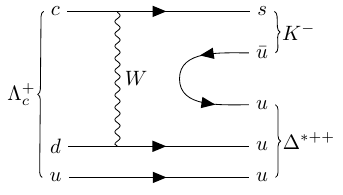}
}
\hspace{1pt}
\subfigure[]
{
\includegraphics[width=0.3\textwidth]{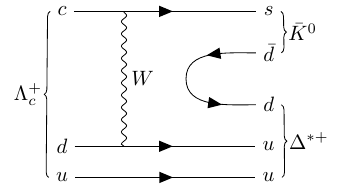}
}
\hspace{1pt}
\subfigure[]
{
\includegraphics[width=0.3\textwidth]{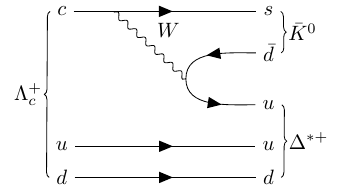}
}
\caption{Feynman diagram for $\lcp \to \Delta(1232)^{++} K^-$ and $\lcp \to \Delta(1232)^{+} \Bar{K}^0$.
}
\label{fig:Feynman_Diagram}
\end{figure*}

\begin{figure*}[thbp]
\centering
\subfigure[]
{
\includegraphics[width=0.45\textwidth]{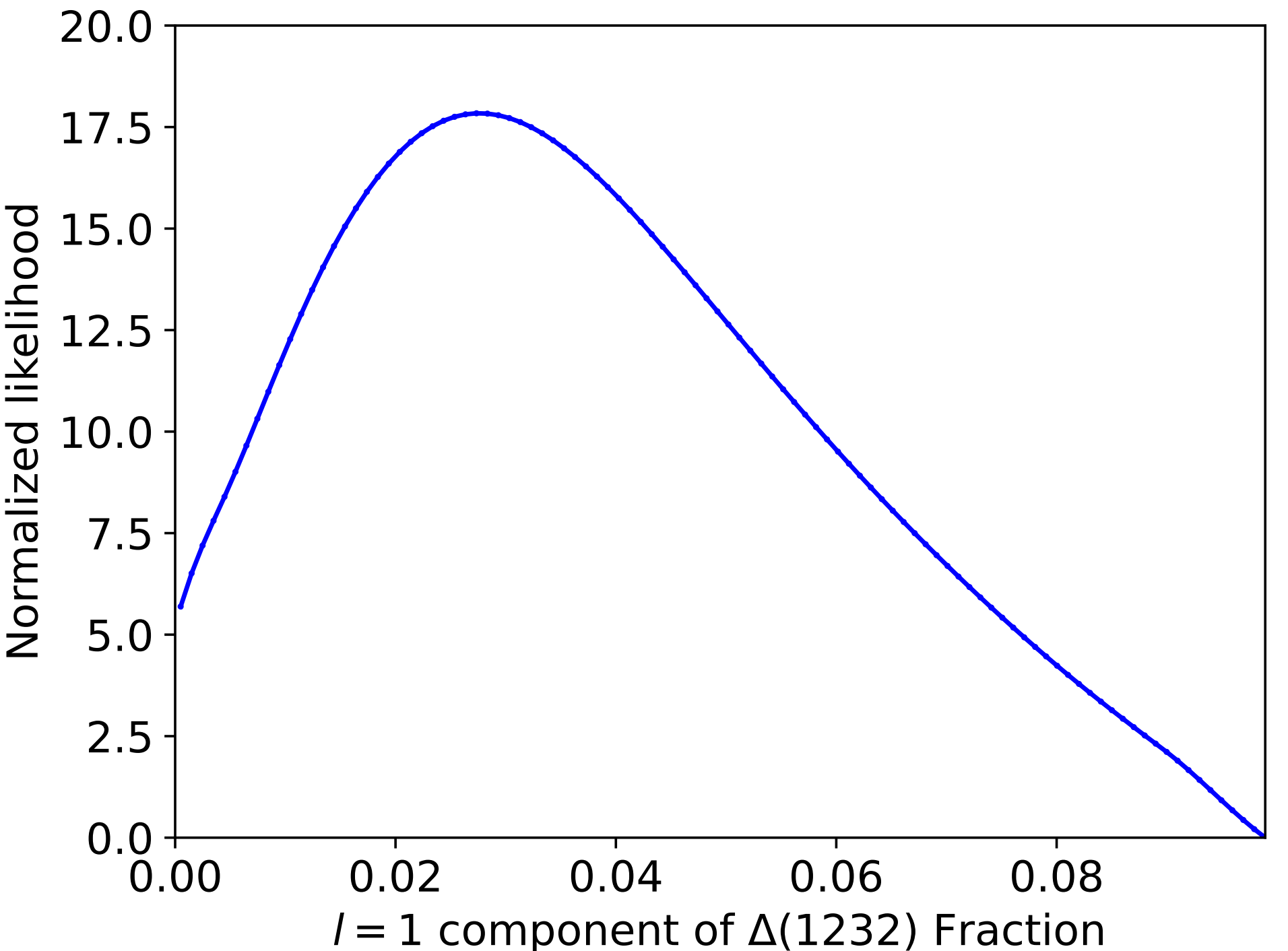}
}
\hspace{1pt}
\subfigure[]
{
\includegraphics[width=0.45\textwidth]{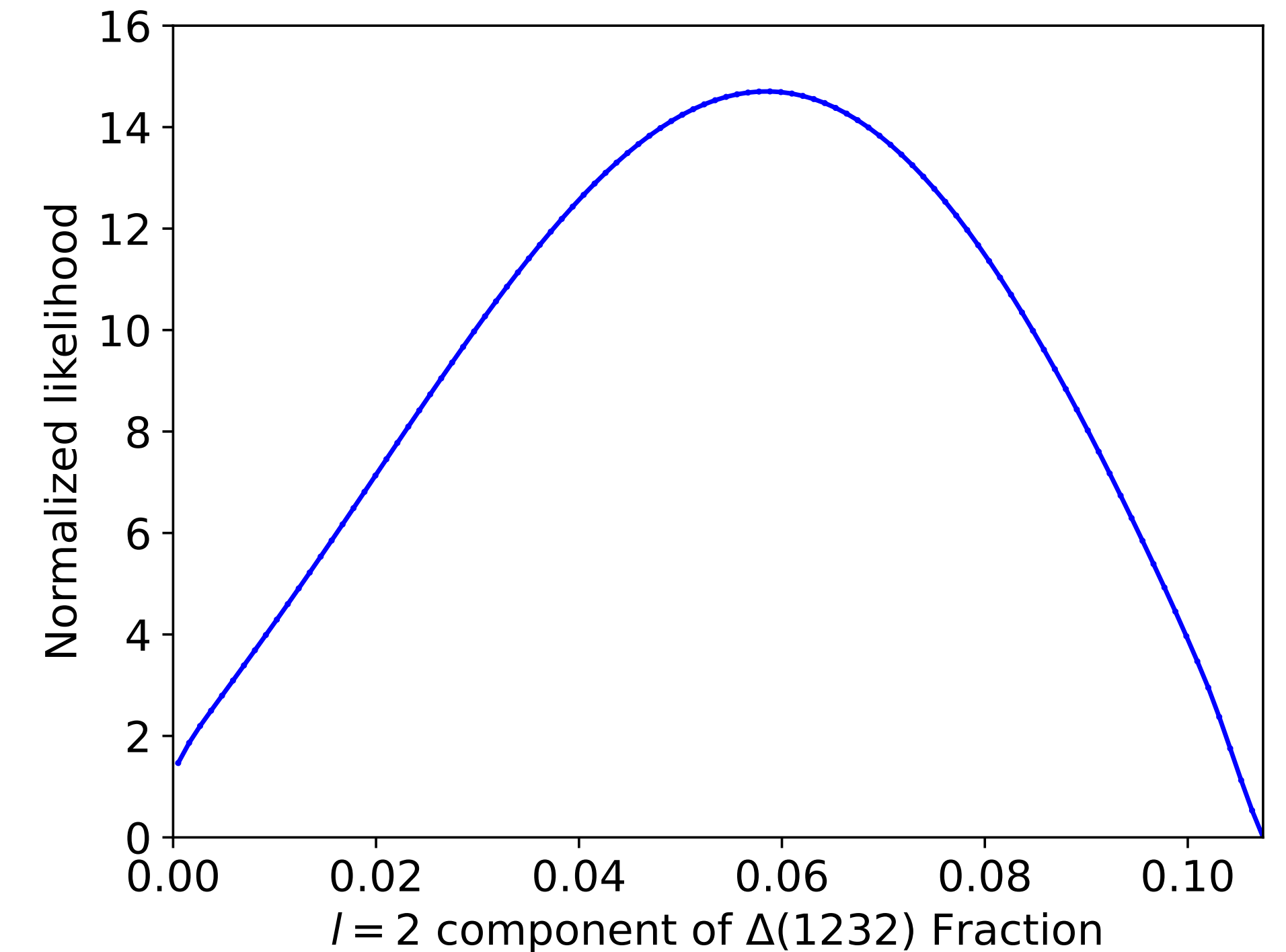}
}
\hspace{1pt}
\subfigure[]
{
\includegraphics[width=0.45\textwidth]{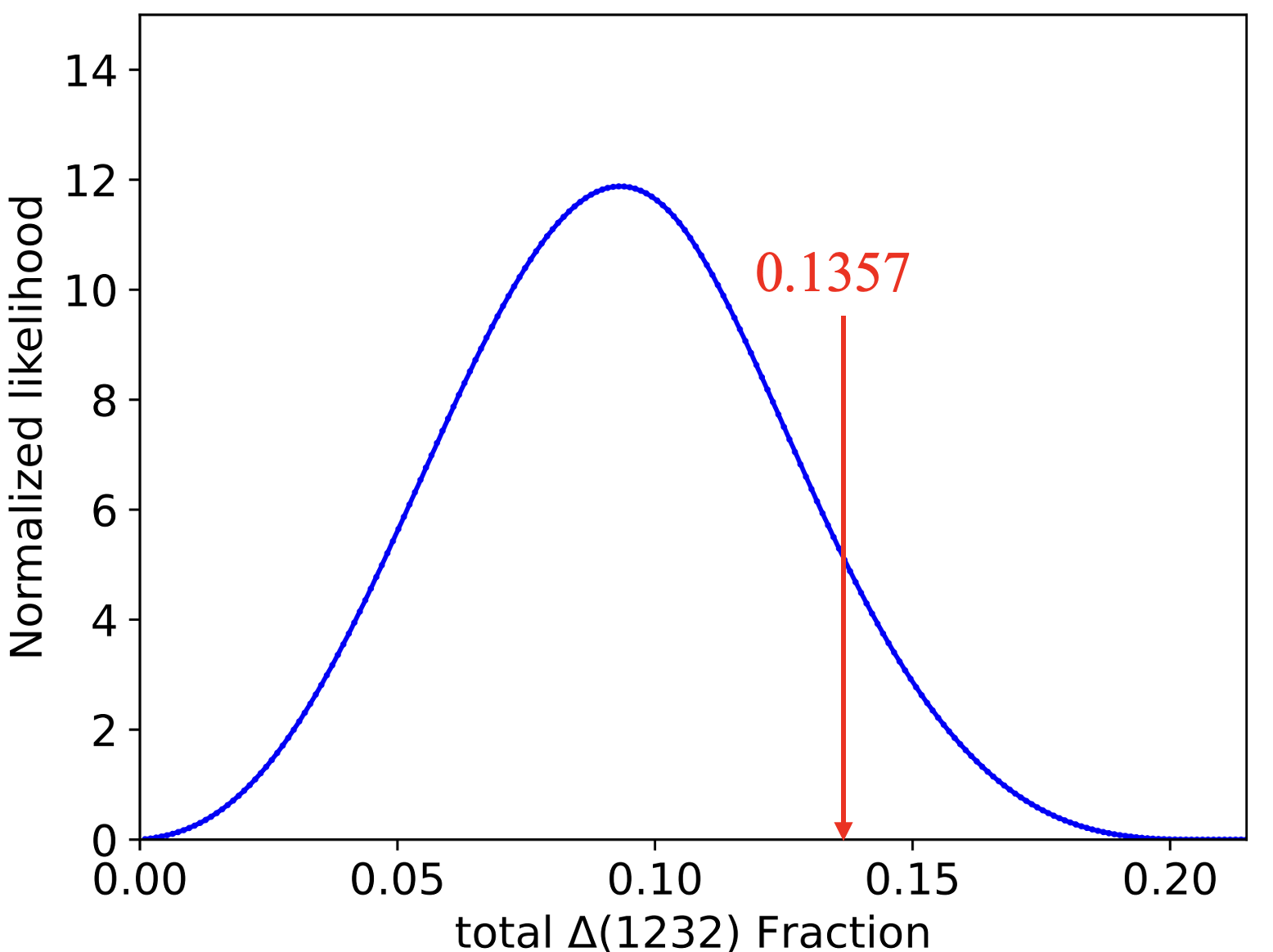}
}
\caption{Normalized likelihood distributions as a function of $l=1$ component fraction (upper left), $l=2$ component fraction (upper right) and total fraction (lower) of $\Delta(1232)^{+}$ in the total rate of $\lcp \to n \Bar{K}^0\pip$. Arrows indicate upper limits at the 90\% confidence level.
}
\label{fig:fraction_scan_upper_limit}
\end{figure*}
For the decay $\lcp \to \Delta(1232)^{++} K^-$, only the $W$-exchange amplitude shown in Fig.~\ref{fig:Feynman_Diagram}(a) contributes at the tree level. In contrast, $\Delta(1232)^{+} \Bar{K}^0$ receives contributions from two tree-level amplitudes: the $W$-exchange and $W$-emission diagrams shown in Fig.~\ref{fig:Feynman_Diagram}(b) and (c). The $W$-exchange topology is identical to that of the $\Delta(1232)^{++} K^-$ mode except a $d\Bar{d}$ pair instead of a $u\Bar{u}$ pair is pulled from the vacuum. Consequently, the $W$-exchange contributes to both $\lcp \to \Delta(1232)^{++} K^-$ and $\lcp \to \Delta(1232)^{+} \Bar{K}^0$, whereas the latter benefits from an additional $W$-emission amplitude. Isospin symmetry implies equal BFs for the subsequent decays $\Delta(1232)^{++}\to p\pip$ and $\Delta(1232)^{+}\to n\pip$; the BF is $(0.9939\pm0.0001)$~\cite{Hunt:2018wqz}.

In our baseline result, the $\Delta(1232)^{+}$ component is not significant, in stark contrast to the large $\Delta(1232)^{++}$ fraction observed in $\lcp \to p K^- \pip$. To determine an upper limit, we perform a likelihood scan for the $\Delta(1232)^{+}$ FFs in the $\lcp \to n \Bar{K}^0\pip$ mode. Figures.~\ref{fig:fraction_scan_upper_limit}(a) and ~\ref{fig:fraction_scan_upper_limit}(b) show the likelihood distributions as a function of the fraction of the $l=1$ and $l=2$ $\Delta(1232)^{+}$ components, respectively. A convolution of these two distributions gives a likelihood profile for the total fraction of $\Delta(1232)^{+}$ in the $\lcp \to n \Bar{K}^0\pip$ mode, which is shown in Fig.~\ref{fig:fraction_scan_upper_limit}(c). Because components with different orbital angular momenta are mutually orthogonal, no interference occurs, so the total fraction equals the sum of the fractions for each orbital‐angular‐momentum component; this sum gives the 90\% confidence-level upper limit on the $\Delta(1232)^{+}$ fraction. However, the distributions of these components are not completely independent: when two orbital‐angular‐momentum components are present simultaneously, the likelihood is lower than that from scanning the components individually. 
Therefore, the upper limit reported is conservative.

We use the integral of the likelihood profile to set upper limits at the 90\% confidence level. Accounting for the BFs of subsequent decays, we obtain $\BR(\lcp \to \Delta(1232)^{+} \Bar{K}^0)<5.08\times 10^{-3}$ at the 90\% confidence level, which is relatively small compared to $\BR(\lcp \to \Delta(1232)^{++} K^-)=(1.79\pm0.09)\%$~\cite{LHCb:2022sck}. We therefore conclude that in $\lcp \to \Delta(1232)^{+} \Bar{K}^0$ decay the $W$-internal-emission and $W$-exchange amplitudes interfere destructively to a large extent, leading to a BF far below that measured for the $\lcp \to \Delta(1232)^{++} K^-$ mode.

\section{Systematic uncertainties}
\label{sec:sys}
\hspace{1.5em} We estimate systematic uncertainties in all observables arising from the following sources: (\romanOne) choice of resonance components; (\romanTwo) fixed fit parameters; (\romanThree) radius parameter; (\romanFour) polarization of $\lcp$; (\romanFive) fit bias from reconstruction and background estimation; and (\romanSix) data-MC difference. Table~\ref{tab:sys_err} lists the sources of systematic uncertainties, which are described below. 

\begin{table}[htbp]
	\centering
	\caption{Absolute systematic uncertainties in the FFs of $\Lambda(1830)^{0}$, $\Lambda(2000)^{0}$, $N(1440)^{+}$, and $N(1650)^{+}$, as well as the corresponding decay asymmetry parameters $\alpha_{\Lambda(1830)^{0}\pip},\alpha_{\Lambda(2000)^{0}\pip}$, $\alpha_{N(1440)^{+}\Ks}$ and $\alpha_{N(1650)^{+}\Ks}$. Total systematic uncertainties are obtained by summing up all contributions in quadrature.\label{tab:sys_err}}
	\vspace{0.2cm}
	$\begin{array}{lcccc}
	\hline \hline 
	& \mathrm{FF}_{\Lambda(1830)^{0}} & \mathrm{FF}_{\Lambda(2000)^{0}} & \mathrm{FF}_{N(1440)^{+}} & \mathrm{FF}_{N(1650)^{+}} \\
	\hline 
	 \mathrm{\romanOne.~Resonance\, components} & 1.90\% & 5.01\% & 6.54\% & 2.95\% \\
	\mathrm{\romanTwo.~Fixed\, fit\, parameters} & 1.40\% & 2.83\% & 3.90\% & 2.70\% \\
	\mathrm{\romanThree.~Radius\, parameter} & 0.27\% & 1.19\% & 1.48\% & 0.60\% \\
	\mathrm{\romanFour.~Polarization\, of\,}\lambdacp & 0.23\% & 0.83\% & 0.42\% & 0.42\% \\
	\mathrm{\romanFive.~Fit\,bias} & 0.35\% & 0.82\% & 0.95\% & 0.36\% \\
	\mathrm{\romanSix.~Data\text{-}MC\, differences} & 0.01\% & 0.02\% & 0.04\% & 0.01\% \\
	\hline
	\rm{Total} & 2.41\% & 5.99\% & 7.83\% & 4.08\% \\
	\hline \hline
	& \alpha_{\Lambda(1830)^{0}\pip} & \alpha_{\Lambda(2000)^{0}\pip} & \alpha_{N(1440)^{+}\Ks} & \alpha_{N(1650)^{+}\Ks} \\
	\hline
	\mathrm{\romanOne.~Resonance\, components} & 0.04 & 0.03 & 0.19 & 0.27\\
	\mathrm{\romanTwo.~Fixed\, parameters} & 0.03 & 0.05 & 0.25 & 0.05\\
	\mathrm{\romanThree.~Radius\, parameter} & 0.02 & 0.12 & 0.15 & 0.02\\
	\mathrm{\romanFour.~Polarization\, of\,}\lambdacp & 0.05 & 0.04 & 0.04 & 0.08\\
	\mathrm{\romanFive.~Fit\,bias} & 0.00 & 0.00 & 0.05 & 0.05\\
	\mathrm{\romanSix.~Data\text{-}MC\, differences} & 0.00 & 0.00 & 0.00 & 0.00\\
	\hline
	\rm{Total} & 0.07 & 0.14 & 0.35 & 0.29 \\
	\hline \hline
	\end{array}$
\end{table}

\begin{itemize}

	\item[\uppercase\expandafter{\romannumeral1}.] 	\textbf{Resonance components}. To estimate the systematic uncertainties from the choice of resonance components, the amplitude fit is repeated by including additional insignificant resonances $\Lambda(1670)^{0}~(2.6\sigma)$; the changes in the fit results among all variations of resonance components are taken as the systematic uncertainties.

	\item[\uppercase\expandafter{\romannumeral2}.] 	\textbf{Fixed parameters}. In the baseline fit, the mass and width parameters of resonances are fixed according to Refs.~\cite{ParticleDataGroup:2022pth,Sarantsev:2019xxm,BESIII:2016tqo,BESIII:2018cnd,Ireland:2019uja}. To estimate the relevant systematic uncertainties, the fixed parameters are varied within $\pm 1\sigma$ and the fit procedure is repeated. The systematic uncertainties are the quadrature sum of the largest variations from each parameter. 

	\item[\uppercase\expandafter{\romannumeral3}.]	\textbf{Radius parameter}. In the baseline fit, the radius parameter $d$ is chosen as $d=1/q_{r}=3.69\,(\mathrm{GeV}/c)^{-1}$ corresponding to the hadron ``scale” parameter $q_{r}=0.2708\,\gevc$. To estimate the systematic uncertainty from this choice, fits are performed setting $q_{r}$ to alternative values derived from Ref.~\cite{BESIII:2019dme}: $q_{r}=0.17\,\gevc$ and $q_{r}=0.37\,\gevc$, which corresponds to the $68\%$ confidence interval. The largest variations are assigned as systematic uncertainties.
	
	\item[\uppercase\expandafter{\romannumeral4}.] \textbf{Polarization of $\mathbf{\lambdacp}$}.
	In the baseline fit, the polarization parameters of $\lambdacp$, i.e. $\alpha_0$ and $\Delta_0$ are fixed according to the previous measurements~\cite{BESIII:2025zbz}. To estimate the systematic uncertainties related to this assumption, the fit is repeated with the polarization parameters varied according to their measured uncertainties. The largest variations are assigned as systematic uncertainties.
	
	\item[\uppercase\expandafter{\romannumeral5}.] 	\textbf{Fit bias}. To estimate the potential bias from the  reconstruction and fit procedure, we perform a full simulation check. We use simulated samples the same size as the data sample. The pull value is calculated as the difference between the fitted and generated, normalized to the fitted statistical uncertainty.
The pull distributions of FFs and decay asymmetries are fit to a Gaussian function with mean ($\mu_{\rm{pull}}$) and standard deviation ($\sigma_{\rm{pull}}$). We correct the baseline fit results $\mu_{\rm{fit}}$ and $\sigma_{\rm{fit}}$ using the fits to the pull distribution:
       \begin{equation}
            \sigma_{\rm{corr}} = \sigma_{\rm{fit}}\times\sigma_{\rm{pull}},~\mathrm{and}~
            \mu_{\rm{corr}} = \mu_{\rm{fit}}-\mu_{\rm{pull}}\times\sigma_{\rm{corr}}.
        \label{equ:correction}
        \end{equation}
Table~\ref{tab:pull_1x_fullsim} gives the values of $\mu_{\rm{pull}}$ and width $\sigma_{\rm{pull}}$. The systematic uncertainty due to these corrections is estimated by propagating the uncertainties in $\mu_{\rm{pull}}$ and $\sigma_{\rm{pull}}$. Here, we do not correct the uncertainties of $\alpha_{\Lambda(1830)^{0}\pip}$ and $\alpha_{\Lambda(2000)^{0}\pip}$, since their uncertainty are determined directly from the NLL scans and do not rely on the Gaussian approximation.

\begin{table}[htbp]
	\centering
	\caption{Pull distribution from the simulation check of FFs and decay asymmetry fit parameters.\label{tab:pull_1x_fullsim}}
	\vspace{0.2cm}
	$\begin{array}{ccccc}
	\hline \hline
	\text { Process } & \multicolumn{2}{c}{\text{FF}} & \multicolumn{2}{c}{\alpha} \\
	\hline
	& \text { Mean } & \text { Width } & \text { Mean } & \text { Width } \\
	\hline
	\lambdacp\to\Lambda(1830)^{0}\pip & \phantom{-}0.40\pm0.07 & 0.94\pm0.05  & / & / \\
	\lambdacp\to\Lambda(2000)^{0}\pip & -0.07\pm0.07 & 0.95\pm0.06 & / & / \\
	\lambdacp\to N(1440)^{+}\Ks & \phantom{-}0.40\pm0.07 & 1.02\pm0.06 & -0.13\pm0.08 & 0.93\pm0.07 \\
	\lambdacp\to N(1650)^{+}\Ks & -0.34\pm0.05 & 1.11\pm0.04 & \phantom{-}0.22\pm0.08 & 0.99\pm0.08 \\
	\hline\hline
	\end{array}$
\end{table}

	\item[\uppercase\expandafter{\romannumeral6}.] 	\textbf{Data-MC differences}. To estimate possible systematic uncertainties due to the difference between the MC-determined efficiency and the experimental one, the effects from tracking and PID of $\pi$ candidates, and the reconstruction of $\Ks$ are considered. The reconstruction efficiency differences between data and MC simulations have been studied by using $J/\psi\to p\bar{p}\pip\pim$ control samples for $\pi$ tracking and PID, and using $J/\psi\to\Ks K^{\pm}\pi^{\mp}$ for $\Ks$ reconstruction. The correction factors $w=\varepsilon_{\mathrm{Data}} / \varepsilon_{\mathrm{MC}}$ are assigned as the weighting factors of the PHSP MC sample and the fit procedure is repeated. Including the weighting factors, Eq.~\eqref{eq:helicty14} becomes
	\begin{equation}
	P(p) = \frac{|\mathcal{A}(p)|^2}{ \sum_{i\in\mathrm{PHSP}} w(p^{i}) |\mathcal{A}(p^{i})|^2 },
	\end{equation}
	where $w(p^{i})$ is the weighting factor for the $i\text{-}\mathrm{th}$ event and the overall constant can be omitted. The resulting variations on the final results are considered as systematic uncertainties.

\end{itemize}

The total systematic uncertainty is obtained by summing all contributions in quadrature. The systematic uncertainties of the BFs and the decay asymmetry are of a size comparable to the corresponding statistical uncertainties.


\section{Summary}
\label{sec:summary}
\hspace{1.5em}

Based on $\ee$ collision samples corresponding to an integrated luminosity of 4.5~$\ifb$ collected with the BESIII detector at c.m.~energies between $4.600\;\gev$ and $4.699\;\gev$, the first amplitude analysis of the charmed baryon hadronic decay $\sigmode$ is performed. Based on the analysis results, the relative BFs for the resonant components are
$$\frac{\BR(\lcp\to\Lambda(1830)^0\pip)\cdot\BR(\Lambda(1830)^0 \to n\Ks)}{\BR(\sigmode)}=( 13.0\pm 3.8 \pm 2.4)\%,$$
$$\frac{\BR(\lcp\to\Lambda(2000)^0\pip)\cdot\BR(\Lambda(2000)^0 \to n\Ks)}{\BR(\sigmode)}=( 29.0 \pm 9.2 \pm 6.0)\%,$$
$$\frac{\BR(\lcp\to N(1440)^+\Ks)\cdot\BR(N(1440)^+ \to n\pip)}{\BR(\sigmode)}=( 21.6 \pm 10.7 \pm 7.8)\%,$$
$$\frac{\BR(\lcp\to N(1650)^+\Ks)\cdot\BR(N(1650)^+ \to n\pip)}{\BR(\sigmode)}=(28.0 \pm 6.0 \pm 4.1)\%.$$
The first uncertainties are statistical and the second are systematic. Using the measured value $\BR(\sigmode)=(1.86\pm 0.08\pm0.04)\%$~\cite{BESIII:2023pia}, the absolute product BFs of the significant intermediate resonant processes are obtained for the first time:
$$\BR(\lcp\to\Lambda(1830)^0\pip)\cdot\BR(\Lambda(1830)^0 \to n\Ks)=(2.4\pm0.7\pm0.5\pm0.1)\times10^{-3},$$
$$\BR(\lcp\to\Lambda(2000)^0\pip)\cdot\BR(\Lambda(2000)^0 \to n\Ks)=(5.4\pm1.7\pm1.1\pm0.3)\times10^{-3},$$
$$\BR(\lcp\to N(1440)^+\Ks)\cdot\BR(N(1440)^+ \to n\pip)=(4.0\pm2.0\pm1.5\pm0.2)\times10^{-3},$$
$$\BR(\lcp\to N(1650)^+\Ks)\cdot\BR(N(1650)^+ \to n\pip)=(5.2\pm1.1\pm0.8\pm0.3)\times10^{-3},$$
where the third uncertainties are due to the input value of $\BR(\sigmode)$. However, contrary to previous theoretical calculation~\cite{Li:2026umb}, neither a narrow peak from $N(1535)^+$ nor a dip due to the $\Lambda(1670)^0$ is observed in the invariant mass spectra of $n\pip$ and $n\Ks$. According to isospin symmetry, the comparison of our results to LHCb results~\cite{LHCb:2022sck} in $\mathcal{B}(\Lambda_{c}^{+} \rightarrow \Lambda(2000)^{0}\pip)\cdot \mathcal{B}(\Lambda(2000)^{0} \rightarrow NK)$ is,
$$
    \mathcal{B}(\Lambda_{c}^{+} \rightarrow \Lambda(2000)^{0}\pip)\cdot \mathcal{B}(\Lambda(2000)^{0} \rightarrow n \Bar{K}^0)= (10.8\pm4.2)\times 10^{-3},
$$
$$
    \mathcal{B}(\Lambda_{c}^{+} \rightarrow \Lambda(2000)^{0}\pip)\cdot \mathcal{B}(\Lambda(2000)^{0} \rightarrow p K^-)= (4.1\pm1.1)\times 10^{-3},
$$
where these two BFs should be equal based on the isospin symmetry. The results are consistent within $1.6\sigma$. Compared to the large fraction of $\Delta(1232)^{++}$ resonance in $\Lambda_c^+\rightarrow pK^- \pi^+$ decay, we determine an upper limit of $\mathcal{B}(\Lambda_{c}^{+} \rightarrow \Delta(1232)^{+}\Bar{K}^{0})$ of $5.08\times10^{-3}$ at 90\% confidence level, indicating significant destructive interference between the two main amplitudes contributing to $\Lambda_{c}^{+} \rightarrow \Delta(1232)^{+}\Bar{K}^{0}$ decay. Other significant resonances observed in the decay mode $\Lambda_c^+\rightarrow pK^- \pi^+$ do not appear to have clear counterparts in this mode, possibly due to the small sample size.
In addition, based on the results of the amplitude analysis, the decay asymmetry parameters for the significant resonant components are determined for the first time:
$$\alpha_{\Lambda(1830)^0\pip}=0.93^{+0.07}_{-0.72}\pm0.07,$$
$$\alpha_{\Lambda(2000)^0\pip}=-0.91^{+0.56}_{-0.09}\pm0.14,$$
$$\alpha_{N(1440)^+\Ks}=-0.13\pm0.43\pm0.35,$$
$$\alpha_{N(1650)^+\Ks}=-0.42\pm0.45\pm0.29.$$

\appendix
\section*{APPENDIX}
\section{Alignment angle calculation}
\label{app:align_angle}
\indent\indent In the helicity formalism, several alignment angles are used in constructing the full amplitude expressed in Eq.~\eqref{eq:helicty13}. Here, we detail the calculations of the alignment angles. The alignment angle is the correction required due to the different definition of $z$-axis in helicity amplitude. Suppose an arbitrary decay $0\to 1+2$, the helicity formula is
\begin{equation}
    \langle p_1, \lambda_1; p_2,\lambda_2 | U | p_0,\lambda_0 \rangle =H_{\lambda_1,\lambda_2} D^{J_0*}_{\lambda_0, \lambda_1 - \lambda_2}(\phi, \theta, 0),
\end{equation}
where $p_{0,1,2}$ are momenta and $\lambda_{0,1,2}$ are helicities of the particles. This means that the evolution operator $U$ can be divided into two parts: a rotation part $R$ 
\begin{equation}
\langle p_0,\lambda_1-\lambda_2|R(\phi, \theta, 0)|p_0,\lambda_0 \rangle = D^{J_0*}_{\lambda_0, \lambda_1 - \lambda_2}(\phi, \theta, 0),
\end{equation}
and an energy-dependent part $H_{\lambda_1,\lambda_2}$. The rotation operator clearly defines $\lambda_1$ and $\lambda_2$ such that the $z$-axis is the direction of $\mathbf{p_1}$ in the rest frame of $p_0$, which corresponds to an Euler rotation $R(\phi, \theta, 0)$. In sequential decay, all rotations from the initial state to the final states should be considered.

The $\sigmode$ process has two decay chains, which are shown in Fig.~\ref{fig:angle}. Following the rotation sequence, we obtain the expressions listed below.
\begin{itemize}
\item For the decay chain $\Lambda_{c}^{+} \rightarrow N^{*+}\Ks$, it is expressed as follows:
\begin{equation}
    R_{N^{*+}} =  
    R(\phi_{n}^{N^{*+}}, \theta_{N^{*+}}, 0)  
    B_z(p_{N^{*+}}) 
    R(\phi_{\Lambda_{c}^{+}}^{1}, \theta_{\Lambda_{c}^{+}}^{1}, 0).
\end{equation}
where the Euler rotation $R(\phi,\theta, 0)$ can be expanded as $R_y(\theta) R_z(\phi)$, $B_z$ denotes boost operation.

\item For $\Lambda_{c}^{+} \rightarrow \Lambda^{*0}\pi^{+}$, it is expressed as
\begin{equation}
    R_{\Lambda^{*0}} =   
    R(\phi_{n}^{\Lambda^{*0}}, \theta_{\Lambda^{*0}}, 0) 
    B_z(p_{\Lambda^{*0}})
    R(\phi_{\Lambda_{c}^{+}}^{2}, \theta_{\Lambda_{c}^{+}}^{2}, 0).
\end{equation}
\end{itemize}

When we take the decay chain $\Lambda_{c}^{+} \rightarrow N^{*+}\Ks$ as a reference, the rotation of alignment is defined as
\begin{equation}
    R_{\Lambda^{*0}}^{\mathrm{align}} = R_{N^{*+}} R_{\Lambda^{*0}}^{-1}.
\end{equation}
Using the 2-dimensional representation of $\mathrm{SU}(2)$ group, the rotation and boost operations can be expressed with:
\begin{equation}
R_z(\phi) = \begin{pmatrix}
e^{-i\frac{\phi}{2}} & 0 \\
0 & e^{i\frac{\phi}{2}}
\end{pmatrix}
,\,
R_y(\theta) = \begin{pmatrix}
\cos\frac{\theta}{2} & -\sin\frac{\theta}{2} \\
\sin\frac{\theta}{2} & \phantom{-}\cos\frac{\theta}{2}
\end{pmatrix}
,\,
B_z(\omega) = \begin{pmatrix}
e^{-\omega/2} & 0 \\
0 & e^{\omega/2}
\end{pmatrix},
\end{equation}
where $\omega = \mathrm{arccosh}{(1/\sqrt{1-(|\mathbf{p}|/E)^2}})$ for a particle with four-momentum $(E,\mathbf{p})$.
We can convert $R^{\mathrm{align}}_{i}$ into several rotations with Euler angles that are the expected alignment angles given by
\begin{equation}
    R_{N^{*+}}^{\mathrm{align}} = B_z(\omega) R_z(\gamma_{p})R_y(\beta_{p})R_z(\alpha_{p}), \ 
    R_{\Lambda^{*0}}^{\mathrm{align}} = B_z(\omega') R_z(\gamma_{p}')R_y(\beta_{p}')R_z(\alpha_{p}').
\end{equation}

\acknowledgments

\hspace{1.5em}
The BESIII Collaboration thanks the staff of BEPCII (https://cstr.cn/31109.02.BEPC) and the IHEP computing center for their strong support. This work is supported in part by National Key R\&D Program of China under Contracts Nos. 2023YFA1606000, 2023YFA1606704, 2025YFA1613900; National Natural Science Foundation of China (NSFC) under Contracts Nos. 11635010, 11935015, 11935016, 11935018, 12025502, 12035009, 12035013, 12061131003, 12192260, 12192261, 12192262, 12192263, 12192264, 12192265, 12221005, 12225509, 12235017, 12342502, 12361141819, 12535005; the Chinese Academy of Sciences (CAS) Large-Scale Scientific Facility Program; the Strategic Priority Research Program of Chinese Academy of Sciences under Contract No. XDA0480600; CAS under Contract No. YSBR-101; 100 Talents Program of CAS; The Institute of Nuclear and Particle Physics (INPAC) and Shanghai Key Laboratory for Particle Physics and Cosmology; Agencia Nacional de Investigaci\'on y Desarrollo de Chile (ANID), Chile under Contract No. ANID CCTVal CIA250027; ERC under Contract No. 758462; German Research Foundation DFG under Contract No. FOR5327; Istituto Nazionale di Fisica Nucleare, Italy; Knut and Alice Wallenberg Foundation under Contracts Nos. 2021.0174, 2021.0299, 2023.0315; Ministry of Development of Turkey under Contract No. DPT2006K-120470; National Research Foundation of Korea under Contract No. RS-2026-25486791; National Science and Technology fund of Mongolia; Polish National Science Centre under Contract No. 2024/53/B/ST2/00975; STFC (United Kingdom); Swedish Research Council under Contract No. 2019.04595; U. S. Department of Energy under Contract No. DE-FG02-05ER41374

\bibliographystyle{JHEP}
\bibliography{mybib}

\newpage
\section*{The BESIII collaboration}
\addcontentsline{toc}{section}{The BESIII collaboration}
M.~Ablikim$^{1}$\BESIIIorcid{0000-0002-3935-619X},
M.~N.~Achasov$^{4,c}$\BESIIIorcid{0000-0002-9400-8622},
P.~Adlarson$^{84}$\BESIIIorcid{0000-0001-6280-3851},
X.~C.~Ai$^{90}$\BESIIIorcid{0000-0003-3856-2415},
C.~S.~Akondi$^{32A,32B}$\BESIIIorcid{0000-0001-6303-5217},
R.~Aliberti$^{40}$\BESIIIorcid{0000-0003-3500-4012},
A.~Amoroso$^{83A,83C}$\BESIIIorcid{0000-0002-3095-8610},
Q.~An$^{79,66,\dagger}$,
Y.~H.~An$^{90}$\BESIIIorcid{0009-0008-3419-0849},
M.~S.~Anderson$^{40}$\BESIIIorcid{0009-0008-1550-2632}
% ...
\\
\vspace{0.2cm}
(BESIII Collaboration)\\
\vspace{0.2cm} {\it
$^{1}$ Institute of High Energy Physics, Beijing 100049, People's Republic of China\\
% ...
$^{4}$ Budker Institute of Nuclear Physics SB RAS (BINP), Novosibirsk 630090, Russia\\
% ...
$^{32}$ INFN Sezione di Ferrara, (A)INFN Sezione di Ferrara, I-44122, Ferrara, Italy; (B)University of Ferrara, I-44122, Ferrara, Italy\\
% ...
$^{40}$ Johannes Gutenberg University of Mainz, Johann-Joachim-Becher-Weg 45, D-55099 Mainz, Germany\\
% ...
$^{66}$ State Key Laboratory of Particle Detection and Electronics, Beijing 100049, Hefei 230026, People's Republic of China\\
% ...
$^{79}$ University of Science and Technology of China, Hefei 230026, People's Republic of China\\
% ...
$^{83}$ University of Turin and INFN, (A)University of Turin, I-10125, Turin, Italy; (B)University of Eastern Piedmont, I-15121, Alessandria, Italy; (C)INFN, I-10125, Turin, Italy\\
$^{84}$ Uppsala University, Box 516, SE-75120 Uppsala, Sweden\\
% ...
$^{90}$ Zhengzhou University, Zhengzhou 450001, People's Republic of China\\

\vspace{0.2cm}
$^{\dagger}$ Deceased\\
% ...
$^{c}$ Also at the Novosibirsk State University, Novosibirsk, 630090, Russia\\
% ...

}

\end{document}